\documentclass[preprint]{aastex}
\usepackage{epsfig}

\shorttitle{FUSE Analysis of EM Cyg} 
\shortauthors{Godon et al.}
\begin{document}
\bibliographystyle{apj}

\title{The White Dwarf in EM Cygni: Beyond The Veil 
\altaffilmark{1}}

\author{Patrick Godon\altaffilmark{2}, Edward M. Sion} 
\affil{Department of Astronomy and Astrophysics
Villanova University,
Villanova, PA 19085;
patrick.godon@villanova.edu; edward.sion@villanova.edu}

\author{Paul E. Barrett}
\affil{United States Naval Observatory, 
Washington, DC 20392;
barrett.paul@usno.navy.mil} 

\author{Albert P. Linnell}
\affil{Department of Astronomy, 
University of Washington, 
Seattle, WA 98195;
linnell@astro.washington.edu} 

\altaffiltext{1}
{Based on observations made with the 
NASA-CNES-CSA Far Ultraviolet Spectroscopic
Explorer. {\it{FUSE}} is operated for NASA by the Johns Hopkins University under
NASA contract NAS5-32985} 
\altaffiltext{2}
{Visiting at the Space Telescope Science Institute, Baltimore, MD 21218;
godon@stsci.edu}

\begin{abstract}

We present a spectral analysis of the Far Ultraviolet Spectroscopic
Explorer ({\it{FUSE}}) spectra of the eclipsing double-line spectroscopic
binary EM Cygni, a Z Cam DN system.  
The {\it FUSE} spectrum, obtained in quiescence, consists of 4 individual
exposures (orbits): two exposures, at orbital phases 
$\phi \sim 0.65$ and $\phi \sim 0.90$, have a lower flux; 
and two exposures, at orbital phases $\phi =0.15$ and $0.45$, 
have a relatively higher flux.
The change of flux level as a function of the orbital phase is consistent 
with the stream material (flowing over and below the disk from the hot 
spot region to smaller radii) partially masking the white dwarf.    
We carry out a spectral analysis of the {\it FUSE} data, obtained
at phase 0.45 (when the flux is maximal), using synthetic
spectra generated with the codes TLUSTY and SYNSPEC.
Using a single white dwarf spectral component,  we obtain a white
dwarf  temperature 
of $40,000 \pm 1000$K, rotating at $100$km/s. 
The white dwarf, or conceivably, the material overflowing the disk
rim, shows suprasolar abundances  of silicon, sulphur and
possibly nitrogen. 
Using a white dwarf+disk composite model, we obtain that the 
white dwarf temperature 
could be even as high as 50,000K, contributing more than 90\% of the FUV
flux, and the disk contributing less than 10\% must have a mass 
accretion rate reaching $10^{-10} M_{\odot}$/yr. The single 
white dwarf model
fits the absorption lines better than the white dwarf+disk model,  but the
white dwarf+disk model fits better the continuum in the shorter wavelengths.  
In both cases, however, we obtain that the white dwarf
temperature is much higher than
previously estimated. We emphasize the importance of modeling the
spectra of EM Cyg around phase $\phi <0.5$, when the white dwarf and disk are
facing the observer, and we suggest that the discrepancy between 
the present analysis and previous  spectral analysis might be due
to the occulting effect of the stream veiling the white dwarf and disk.       

\end{abstract}


\section{Introduction}
\subsection{Dwarf Novae}

Dwarf Novae (DNe) are a class of non-magnetic cataclysmic variables (CVs) 
in which a white dwarf (WD) accretes matter from a main-sequence 
star (the {\it secondary}) filling its Roche lobe 
by means of an accretion disk.  
DN systems  are characterized mainly by their brightness variations, due  
to a periodic change in the accretion rate. 
Ongoing accretion at a low rate (quiescence) is interrupted
every few weeks to months by intense accretion (outburst) of days to
weeks which increases the luminosity of the system
\citep{hac93,war95}. 
This outburst is believed to be due to an instability
in the disk (the disk instability model - DIM \citep{can98}), 
and it is known as a {\it dwarf nova} accretion event 
(or dwarf nova outburst). 

Due to their long term accretion, the WDs
in DNe have temperatures $\sim 15,000-50,000$K (higher than field 
WDs), and therefore peak in the far-ultraviolet (FUV).  
The long periods of quiescence allow for long exposure observations 
($\sim$hrs) covering all the orbital phases  
(this is especially important for systems with a high inclination,   
as the WD can be eclipsed or masked at given orbital phases).  
Because of that, DNe, unlike other CVs, are ideal candidates to
study the accreting WD, as during quiescence the WD is exposed in the FUV for
long periods of time. 
FUV spectroscopic observations are used to determine the
WD parameters such as temperature, gravity, rotational velocity 
and chemical abundances.
The disk  mass accretion rate of many systems 
can also be deduced accurately at given epochs
of outburst or near outburst using spectral fitting
techniques. 
In addition, DNe offer a fairly reliable estimate of
their distances via the absolute magnitude at maximum versus orbital
period relation \citep{war95,har04}. 
As a consequence,  DNe are ideal candidates to observe, as one can 
more easily derive their system parameters, which in turn can be used to 
test the theories of CV evolution.       

Dwarf nova systems are further divided into sub-types: 
U Gem systems, SU UMa systems, and Z Cam systems. 
The U Gem systems are the typical DNe, i.e. with normal DN outbursts; 
the SU UMas exhibit both normal DN outbursts, and
superoutbursts, which are both longer in duration and higher in 
luminosity than normal DN outbursts; 
and the Z Cam systems have standstills where they remain in a semi-outburst
state for a long time \citep{rit03}. 
The binary orbital period in CV  
systems ranges from a fraction of an hour to about a day; however, 
there is a gap in the orbital period
between 2 and 3 hours where almost no CV systems are found: 
the {\it period gap}. U Gem and Z Cam
DN systems are found above the period gap, while the SU UMa DN systems
are found below the period gap.  

In this paper we present the FUV spectral analysis of the eclipsing
Z Cam system EM Cygni observed during quiescence with {\it FUSE}.

\subsection{EM Cygni System Parameters} 

First suspected to be related to CVs by \citet{bur53}, EM Cyg  
is a bright, double-lined spectroscopic binary variable \citep{kra64}. 
Its eclipses were discovered by \citet{mum69} 
and the orbital period was found to be almost 7h (0.29090912 day).  
Its period was expected to decrease (by $\sim10^{-11}-10^{-10}$ 
days/cycle; \citet{pri75,mum80}), however it   
has been found to be constant \citep{beu84,csi08} with an upper limit  
$\dot{P}/P < 2.3 \times 10^{-12}$/days. 
Its outburst cycle is of the order of 25 days
and rapid coherent oscillations are detected during the late
stages of outbursts
with periods of the order of 14.6-16.5 seconds (see 
\citet{sti82} and references therein).    

EM Cyg is so far the only known Z Cam subtype eclipsing DN system. 
The eclipses are important as they allow, in theory at least, 
for a detailed study of the system. 
In addition, because EM Cyg is an eclipsing system as well 
as a double-lined
spectroscopic binary, it has been extensively studied 
(e.g. \citet{rob74,nev78,bai81,jam81,pat81,sto81,szk81}; 
see also \citet{sti82}). 
However, the determination of the stellar masses of the system  
by \citet{rob74} and \citet{sto81} was effected by 
the presence of a third star, {\it Arkadash}
\footnote{ 
The third star was named {\it Arkada\c{s}} (``friend'' in Turkish)
by \citet{wel05}, in memory of Janet (Hannula) Aky\"uz Mattei (1943-2004). 
Janet Mattei determined an outburst cycle of 25 days for EM Cyg, 
and provided AAVSO data and information on the state of EM Cyg 
many times (e.g. \citet{sto81,sti82,nor00}), 
as she did for many variable stars on countless occasions.
}, 
along the line of sight \citep{nor00}. This third star is not
related to the system.  
The most recent study \citep{wel07} indicates a mass ratio $q=0.77$ 
and a WD mass $M_{wd}=1M_{\odot}$. 
The distance found by \citet{bai81,jam81}, $d\sim$320-350pc, 
could also be affected by the contribution of Arkadash in the red light, 
and \citet{wel05} pointed out that $d$ could 
be as large as $\sim$450-500pc. However, using the maximum magnitude
at outburst versus orbital period relation \citep{war95,har04},
one finds a distance of $\sim 400$pc \citep{win03}.

A recent broad-band photometric study of the system \citep{spo03,spo05}
shows the magnitude of EM Cyg varies between 
V=13.3 at minimum and V=12.5 at maximum,  
it reaches V=14.4 during the eclipse, and V=12.9 in standstill    
(note, however, that data from the AAVSO indicates it can reach V$\sim$12.0).  
EM Cyg is located at a
very low Galactic latitude ($b=4.28^{\circ}$), 
where the Galactic reddening reaches almost 0.5. 
However, the reddening towards EM Cyg is only  $0.05\pm 0.03$ 
\citep{ver87,lad91,bru94}. 
The system has an inclination $i$ of {\it only} $67\pm 2$deg \citep{nor00},
however, because of its large disk and secondary star, 
the disk is partially eclipsed, while the WD itself is not eclipsed. 
The geometrical configuration
of the disk, WD and secondary of the system is shown in Figure 1 
at phase $\phi=0/1.0$.  
All the system parameters are recapitulated in Table 1.  

\subsection{Previous Spectral Analysis of EM Cyg} 
 
EM Cyg was observed in quiescence with {\it FUSE} 
by \citet{wel05}, who
carried out a spectral fit after combining  
the {\it FUSE} spectrum with an {\it IUE} spectrum. 
Their spectral fit consists of a disk with
$\dot{M} = 2.5 \times 10^{-12} M_{\odot}$/yr and a $0.88M_{\odot}$ WD with
$T \approx 10-20,000$K, assuming a distance of 350pc.  
The model did not fit the flux in the short 
wavelengths of {\it FUSE} which is indicative of a much higher temperature.
However at $T>20,000$K the model spectrum became too blue and the  fit
was noticeably worse in the {\it IUE}  range. 

\citet{win03} carried out a spectral fit to the {\it IUE} spectrum 
SWP 08088 of EM Cyg  taken in quiescence 
(with a flux of $\sim 3 \times 10^{-14}$ergs$~$cm$^{-2}$\AA$^{-1}$) 
and obtained that the best fit is a 
disk with  $\dot{M} = 5 \times 10^{-11} M_{\odot}$/yr and a 
WD of $1M_{\odot}$ with
$T < 24,000$K (contributing only 8\% of the flux) with a distance of 210pc
and $i=75^{\circ}$.  
Later on, \citet{urb06} analyzed the same {\it IUE} spectrum 
and obtained the same results except a distance of 350pc, 
when dereddening the spectrum assuming E(B-V)=0.05. 
\citet{ham07} analyzed the {\it IUE} spectrum SWP 07297 of EM Cyg 
in outburst assuming d=350pc, $i=75^{\circ}$, and obtained that the
flux comes mostly from a disk with $\dot{M}=7 \times 10^{-9}M_{\odot}$/year. 

In the spectral analysis of \citet{win03} and \citet{urb06} 
only the {\it IUE} spectra of EM Cyg in quiescence was modeled.  
While in the work of \citet{wel05} the fitting of the 
short wavelength range of {\it FUSE} ($<1000$\AA ) was not carried out
satisfactorily (as pointed out by \citet{wel05} themselves). 
It is the purpose
of this work to carry out a spectral fit to the {\it FUSE}
spectrum of EM Cyg in quiescence including the short wavelength range, 
and to obtain the physical parameters of the exposed WD.

We give particular importance to the orbital 
phase at which the individual {\it FUSE} exposures were obtained.
We show that at particular orbital phases the WD might be veiled  by the
stream from the first Lagrangian point overflowing the disk. 
We remark that previous estimates might have modeled a flux attenuated 
by as much as 75\%, and consequently obtained a low WD temperature.
We model here the {\it FUSE} spectrum of the WD when veiling 
is minimum and obtain a temperature $T\sim 40,000$K and possibly as high
as 50,000K.  

In the next section we present the four individual {\it FUSE} 
exposures and compare them with the {\it IUE} spectra
taken in quiescence; in section 3 we review the stream-disk interaction 
and in particular we check the possibility of the stream overflowing the disk; 
in section 4 we describe the synthetic
stellar spectral code as well as our fitting technique; the results
are presented in section 5; and a discussion with our conclusions 
are given in the last section.

\section{Observations of EM Cyg} 

\subsection{The {\it FUSE} Spectrum}

The {\it{FUSE}} data of EM Cyg were obtained on 2002 Sep 5 (HJD2452522-3,
when the system was in a low state, see Figure 2) 
through the 30"x30" LWRS Large Square Aperture in TIME TAG mode. 
The data were processed with the latest and final version of 
CalFUSE (v3.2.0; \citet{dix07}).
We process the {\it{FUSE}} data as in \citet{god08}, 
and details of the procedure can be found there.  

The {\it FUSE} spectrum of EM Cyg consists of 4 individual exposures
(one per orbit). For the purpose of line identifications we present the
averaged spectrum in Figure 3. 
We identify many ISM molecular hydrogen absorption lines as
well as some of the metal lines usually seen in the {\it FUSE}
spectra of accreting WDs.  
The metal lines we identify are as follows:  \\  
- the S\,{\sc iii} (1021 \AA ),  
S\,{\sc iv} (1062 \& 1073 \AA ),  
S\,{\sc vi} (945 \AA ) lines   
and O\,{\sc vi} doublet, all indicative of a higher temperature
$T \sim 35,000$K and much higher;  \\  
- the Si\,{\sc iii} (1114 \AA\ ),  
Si\,{\sc iv} (1122 \AA\ ), 
and N\,{\sc ii} (1085 \AA\ ) lines, sometimes indicative of a lower
temperature $T\sim 25,000$K, but not inconsistent with a higher
temperature; \\  
- the C\,{\sc iii} (1175 \AA ) absorption line,  
present at all WD temperatures, as long as the carbon abundance 
is a fraction of its solar value. \\  
The C\,{\sc ii} (1010\AA \& 1066\AA ) absorption lines, seen 
at lower temperatures ($T\sim 25-30,000$K), are not  
detected here (though the C\,{\sc ii} (1010\AA ) could be contaminated
with ISM molecular absorption).  
There are some broad emission features  from 
C\,{\sc iii} (977 \AA ), and the O\,{\sc vi}  doublet. 
The C\,{\sc iii} (1175 \AA ) presents a P-Cygni profile.
The sharper absorption lines are from the 
ISM and possibly also from circumbinary material. It is likely that the
nitrogen absorption lines (N\,{\sc i} \& N\,{\sc ii}) are 
contaminated with terrestrial absorption. All these lines are marked
and identified in Figure 3. There are a few additional absorption
features/lines that we cannot identify, mainly in the longer wavelengths
(these lines are not labeled in the lower panel of Figure 3).  
The sharp emission lines are from air glow (geo- and helio-coronal 
in origin).
The broad Ly$\beta$ absorption feature is not clearly seen, pointing
to a high temperature, some broad oxygen doublet emission contamination, 
and/or possibly a rotational velocity broadening from a disk component. 
In the short wavelengths ($< 950$\AA ) the flux does not go to zero
and the continuum there is about 1/3 of the flux at longer 
wavelengths ($>1000$\AA ), indicative of a high temperature and
consistent with the presence of higher ionization metal lines.

As mentioned, this {\it FUSE} spectrum consists of 4 individual
exposures taken during 4 consecutive orbits of the {\it FUSE} 
telescope. 
The four exposures correspond to the binary phases $\phi \sim$0.45, 
0.65, 0.90, and 0.15,
where $\phi=0$ corresponds to the eclipse
\citep{wel05,wel07}.  
The observations log for the individual {\it FUSE} exposures
is given in Table 2. 
Using an analysis of the lines in the four individual {\it FUSE} 
exposures, \citet{wel05,wel07} explicitly showed how
the radial velocity changes as a function of the orbital phase. 
However, they did not consider the variation of the continuum flux level from
exposure to exposure, i.e. the variation of the FUV flux as a function
of the orbital phase. In Figure 4 we show how the flux varies from
exposure to exposure, and this change is also registered in the
visual magnitude of the system as shown in data from the AAVSO in
Figure 2. We find that at phases 0.15 \& 0.45 (exposures 4 \& 1) 
the {\it FUSE} continuum flux is larger (by up to a factor three)
than at phases 0.65 \& 0.90 (exposures 2 \& 3).   

In order to understand the variation of the FUV flux,   
in Figures 5 we show the binary configuration at the times the 4 
{\it FUSE} exposures were obtained. From the orbital configuration
we see that the flux in exposures 2 \& 3 is attenuated when the
hot spot region  and the stream overflow region are in front of the
observer.   
In the next subsection we carry out an examination
of the IUE spectra of EM Cyg, and we find that they too present
some flux variation as a function of the orbital phase.  
We suspect that the FUV flux is possibly attenuated by   
absorption from the stream material (originating at L1) overflowing 
the disk edge and reaching a height above (and below) the
disk of the order of $z/r \sim 0.4$
We further check  this suggestion  in section 3.  

\subsection{The {\it IUE} Spectra}

Since the {\it FUSE} flux varies as a function of the orbital phase
it is interesting to compare it with other spectra obtained in quiescence;
this will also help us interpret results from previous spectral analysis.  
The existing archival {\it HST}/STIS 
spectra of EM Cyg were taken during
outburst, just a week before the acquisition of the
quiescent {\it FUSE} spectrum (see Figure 2), 
and the three existing {\it HUT} spectra were also taken in outburst. 
Because of that we disregard the STIS and HUT spectra.
However, some of the existing {\it IUE} 
spectra were taken in quiescence, and later were modeled by 
\citet{win03,wel05,urb06}. We present and discuss these spectra here.  

In Figure 6 we compare three {\it IUE} SWP spectra
taken in quiescence together with the {\it FUSE} exposures 
in the spectral range $\sim 900-1,400$\AA\ .   
Since {\it FUSE} exposures 2 and 3 have about the same flux, we 
combine them together, and the same for exposures 1 and 4. 
We see that SWP 08088 (taken in 1980) has a much lower flux, 
similar to exposures 2 and 3.
SWP 17591 (taken in 1982) has a flux level comparable with exposures 
1 and 4, while SWP 17592 (taken 90min after SWP 17591) 
has a higher flux level. 

The typical quiescent $V$ of the system 
is 13.5 but it has been observed as faint
as 14.5 in quiescence (possibly at eclipse \citep{spo03,spo05}).  
From validated AAVSO data, it appears that, at the time of the
{\it FUSE} observations, the system had a magnitude of  
$V \sim 13.1-13.9$ (Figure 2), namely  
EM Cyg wasn't in its lowest possible quiescent 
state at the time of the {\it FUSE} observations.  
This is clearly seen in Figure 2, where the previous  
quiescent state on JD2452505 is about 0.5mag fainter than the 
quiescent state during which the {\it FUSE} observations (JD2452522) 
were carried out. 
The IUE SWP 17591 and SWP 17592 spectra were obtained during 
a brighter state, when EM Cyg was on the rise to
outburst, its magnitude around 13.2-13.5 (but still lower than
a standstill with $V\sim 12.5-13.5$). 
The {\it IUE} SWP 08088 spectrum was at $V\sim 13.6-13.7$ \citep{szk81}.  

The three {\it IUE} SWP spectra are complemented with three
LWR spectra. 
All the quiescent (LWR and SWP)
{\it IUE} spectra are listed in Table 2 together with their
corresponding orbital phase. 
The increase in flux between SWP 17591 and
SWP 17592 over a time scale of 90min is most probably due 
to orbital variation (as in {\it FUSE}), and the matching LWR
spectra 13862 and 13863 do not show any variation. 
As to the SWP 08088 and (matching) LWR 07056 spectra, their lower flux is  
due to a deeper quiescence and to orbital variation too.  

Because of these changes in flux level (due to both orbital
variation and intrinsic changes in the source) combining the
{\it FUSE} and {\it IUE} spectra together is tricky and this might
explain the difficulty \citet{wel05} were having to match both the
short wavelengths of {\it FUSE} and the longer wavelengths
of {\it IUE} in the same fit. 
As a consequence we do not combine the {\it FUSE} spectrum
of EM Cyg with any {\it IUE} spectrum. 
The shorter wavelength range as well as the higher
S/N of {\it FUSE} over {\it IUE} makes the {\it FUSE} spectrum of
EM Cyg ideal for determining the temperature of the accreting WD.

From the {\it FUSE} \& {\it IUE} spectra of EM Cyg, 
we found that the FUV flux between $\sim$900\AA\ and 2000\AA\ 
is attenuated around phase $\phi \sim 0.6-0.9$. 
Since the {\it FUSE} spectrum presents many deep absorption lines,  
it is tempting to suggest the possibility of an absorbing component moving 
in front of the observer around phase $\sim 0.75$, as the binary
rotates. We further investigate this possibility in the next section.  

\section{The Stream-Disk Interaction}

The only components of the system that could be in the field of view
of the WD/inner disk around $\phi \sim 0.6-0.9$ are the region downstream the
hot spot and the stream possibly overflowing disk at the hot spot.     
In order to check this possibility, in this section,  we consider 
previous studies of the stream-disk
interaction, and also present a simple ballistic trajectory
simulation using the binary parameters of EM Cyg. We show that
for the binary parameters of EM Cyg, stream material could be found
high enough above the disk of the plane, around phase $\phi \sim 0.6-0.9$
to actually veil the WD and inner disk.  

Pioneering analytical studies by \citet{lub76,lub89} have shown how part of
the stream material possibly flows over and under the disk 
and falls back onto the disk at smaller
radii at an angle $\phi_{impact} \approx 140-150^{\circ}$ 
($\phi_{impact}$ is measured counter
clockwise from the line joining the two stars), corresponding to
an orbital phase $\phi \sim 0.6$. 
Observationally, Doppler tomography \citep{mar85} has been used for 
a quarter of a century to confirm the stream disk interaction and the
impact at smaller radii in some system (e.g. Z Cha). 
In \citet{lub89} some of the
stream matter was flowing over the disk at a scale height 
$\sim 0.01a$, where $a$ is the binary separation. For a disk of 
diameter $R_{disk}=0.3a$, this corresponds to a scale height of
$z/r \approx 0.03$ at the edge of the disk. Namely, in these analytical
studies, it was obtained that stream material overflows the
disk rim (at the hot spot) and continue toward smaller 
radii while basically {\it grazing} the disk's surfaces (at
a few vertical scale lengths - a few $\sim H$).   
Therefore, based on these studies alone, in order for the material
to veil the WD, the inclination of the system has to be 
$\cos{i} \approx 0.03 \rightarrow i \approx 88^{\circ}$ or larger:
the system has to be nearly edge-on.  

However, more recent full 3D hydrodynamical simulations,  
e.g. \citep{blo98,kun01,bis03},  
have followed in detail the interaction of the material
streaming from the first Lagrangian point (L1) with the disk.    
The main result from these simulations is that the 
stream does not simply grazed the disk's surfaces, but 
matter is actually deflected vertically from the hot spot and flows in a more
or less diffuse stream to inner parts of the disk, hitting the disk
surface close to the {\it circularization radius} $r_c$ (see \citet{lub75}
for details on $r_c$) at orbital phase 0.5. 
In these simulations, the stream material deflected vertically
at the hot spot region can reach an altitude $z \sim r$.  
This stream deflection 
is believed to cause X-ray absorption in CVs (and LMXBs) around orbital
phase 0.7, if the inclination is at least $65^{\circ}$ \citep{kun01}.    
These simulations also explain streams of material perpendicular to the orbital
plane as observed in Algol-type binaries (e.g. U CrB; \citet{aga09} and
references therein).  
These results imply that stream material can be deflected vertically, 
and reaches a height $z/r=\cos{i}=\cos{65^{\circ}}=0.42$ and lands
onto the disk at $\phi_{impact}=180^{\circ}$ (corresponding to 
an orbital phase 0.5) \citep{kun01}. 

We suggest the same might be happening in EM Cyg, and the decreased 
flux around $\phi \sim 0.7$ might be caused by the deflected
stream flow. In order to verify the plausibility of this scenario,
we simulate the stream flow from the L1 point using the ballistic
trajectory approximation, and we deflect the ballistic particle at the
hot spot by an angle $\theta$ (measure from the plane of the disk).  
Since the stream flow consists of a infinite
number (in the continuity limit) of particles (and since the stream
has a finite thickness extending vertically), the ballistic trajectory
is solved for all possible value of $\theta$ (from $-90^{\circ}$ 
to $+90^{\circ}$). A maximum
altitude is reached for a critical angle of deflection
$\theta =\theta_{Cri}$. This maximum altitudes
depends on the values of the system parameters:
veiling will not occur for every mass ratio, disk radius
and inclination 
(the full details of the simulations as well as results for the
entire space parameters of CVs are given elsewhere \citep{god09}). 
Figure 8 depicts such a simulation for EM Cyg, where we show with
arrows the 4 {\it FUSE} exposures. The mass ratio is 0.77 and
$R_{disk}=0.3a$. The stream is shown in red and the
blue area represents that portion of the trajectory  for which 
the material reaches $z/r \ge 0.4$, corresponding to veiling at 
$i=67^{\circ}$. 
Here the stream is deflected at the hot spot with 
an angle $\theta_{Cri} \approx 55^{\circ}$
above (and below) the plane of the disk. This is the angle for which
the altitude of the ballistic particle reaches a maximum. For a smaller,
or {\it larger}, 
deflection angle $\theta$, the blue region shrinks.  For a smaller
disk radius ($R_{disk}<0.3$) the maximum elevation above the disk increases, 
however, the hot spot moves counterclockwise towards $\phi \sim 0.75$ 
and as a consequence exposure (3) is not as veiled as for 
$R_{disk}=0.3a$.  This points to the 
importance of the large disk in EM Cyg for deflecting material
off the disk plane around $\phi \sim 0.6-0.9$.  
This picture is consistent with the full 3D hydrodynamical  simulations of 
\citet{kun01}. For different system parameters 
(inclination, mass ratio and disk radius) the deflected material
does not especially veil the WD, as the maximum vertical elevation of the 
ballistic particle does not always reach $z/r=\cos{i}$. 

From this specific ballistic trajectory simulation with the system
parameters of EM Cyg, we conclude that the lower flux in the {\it FUSE} 
exposures 2 and 3
can be accounted for with stream material being deflected 
vertically at the hot spot forming a veil in front of the WD
between orbital phases $\sim 0.65$ and $\sim 0.90$ (for an inclination
of $67^{\circ}$). The increase in flux between the {\it IUE} spectra 
SWP 17591 ($\phi\sim0.75$) and
SWP 17592 ($\phi\sim0.97$) over a time scale of 90min can also be explained by  
veiling of the WD/inner disk by the stream overflow around 
phase 0.75 (keeping in mind that the WD/inner disk are not eclipsed
around $\phi \sim 1.0$).

Furthermore, the stream material overflowing the disk edge lands onto
the disk at a smaller radius corresponding to orbital phase of up to 
$\phi \sim 0.5$, 
and produces a bright spot there \citep{mar85,lub89}.  
Together with the hot spot, this contributes to additional flux in the longer
wavelengths (LWR spectra).  
The emission from the hot spot and point of impact is possibly 
seen in 
the LWR spectra (13862, 13863) shown in Figure 7 as a plateau in the 
longer wavelengths
($\lambda \sim 2250-3000$\AA ), where one rather 
expects the flux to decrease there. This flux
in the longer wavelengths  was already associated with emission from 
the hot spot by \citet{szk81}.  
Though we have no way to confirm that the plateau in the longer wavelengths
of {\it IUE} is due in part to the impact point of the stream overflow, 
it is, however, consistent with this possibility.  
The plateau seen in the LWR spectra, is also contributing
significantly to the longer wavelengths of the SWP spectra. 
This is especially true for the SWP 08088 spectrum.  
These are additional reasons not to model the {\it IUE} spectra, 
nor to combine them with the existing {\it FUSE} spectrum, as the
hot spot and point of impact contribute flux to the longer wavelengths
of the {\it IUE} SWP spectra.

\section{Synthetic Spectral Modeling} 

\subsection{Modeling the Stellar Spectrum} 
 
We create model spectra for
high-gravity stellar atmospheres using the codes 
TLUSTY and SYNSPEC\footnote{
http://nova.astro.umd.edu; TLUSTY version 200, SYNSPEC version 48} 
\citep{hub88,hub95}. 
We generate photospheric models for a $1.0M_{\odot}$ WD with effective
temperatures ranging from 12,000K to 75,000K in increments of 
about 10 percent (e.g. 1,000K for T$\approx$15,000K and 5,000K for 
T$\approx$70,000K).
We vary the
stellar rotational velocity $V_{rot} sin(i)$ from $50$km$~$s$^{-1}$
to $500$km$~$s$^{-1}$ in steps of $10-50$km$~$s$^{-1}$. 

We also use disk models from the grid of models generated by 
\citet{wad98}. We select their models with a WD mass $M=1M_{\odot}$ 
and inclinations $i=60^{\circ}$ \& $75^{\circ}$. These models have
solar abundances, but because of the velocity broadening the
absorption lines are completely smoothed out, such that the disk models
($i\sim 60^{\circ}-75^{\circ}$) 
are not sensitive to the chemical abundances.   These disk models
are either used as such or in combination with WD models we generate.

After having generated grids of models,  
we use FIT \citep{numrec}, a $\chi^2$ minimization routine,
to compute the reduced $\chi^{2}_{\nu}$ 
($\chi^2$ per number of degrees of freedom) 
and scale factor values for each model fit.  
Before carrying out each synthetic spectral fit of the spectrum,
we masked portions of the spectrum with strong emission lines,
and air glow.
While we use a $\chi^2$ minimization technique, we do not 
blindly select the least $\chi^2$ models, but we also examine the models 
that best fit some of the features such as absorption lines. 

The WD rotation ($V_{rot} sin(i)$) rate is determined by fitting the
WD model to the spectrum while paying careful attention to the 
line profiles.
We did not carry out separate fits to individual lines but
rather tried to fit the lines and continuum in the same fit.
We first fit solar abundance models and
only then we change the abundances of the species.

\subsection{Modeling the ISM Hydrogen Absorption Lines} 

We identify ISM   lines in Figure 3 to
avoid confusing them with the WD lines. 
The ISM lines are deep and broad and we decided 
to model them, especially since some of the WD lines (such as 
S\,{\sc iv} $\lambda  \lambda$1062.6 \& 1073)  are located 
at almost the same wavelengths.   

We model the ISM hydrogen absorption lines to assess
the atomic and molecular column densities. This enables us to 
improve the WD spectral fit. 
The ISM models (transition tables) are generated using a program developed by 
P.E. Barrett.     
This program uses a custom spectral fitting package to estimate the temperature
and density of the interstellar absorption lines of atomic and
molecular hydrogen.  The ISM model assumes that the temperature, bulk
velocity, and turbulent velocity of the medium are the same for all
atomic and molecular species, whereas the densities of atomic and
molecular hydrogen, and the ratios of deuterium to hydrogen and metals
(including helium) to hydrogen can be adjusted independently. The
model uses atomic data of \citet{mor00,mor03} and molecular data of
\citet{abg00}. The optical depth calculations of molecular
hydrogen have been checked against those of \citet{mcc03}.

The ratios of metals to hydrogen and deuterium to
hydrogen are fixed at 0 and $2 \times 10^{-5}$, respectively,
because of the low
signal-to-noise ratio data.  The wings of the atomic lines are used to
estimate the density of atomic hydrogen and the depth of the
unsaturated molecular lines for molecular hydrogen.  The temperature
and turbulent velocity of the medium are primarily determined from the
lines of molecular hydrogen when the ISM temperatures are $< 250$K.

The ISM absorption features are best  modeled and displayed when the
theoretical ISM model (transmission values) is combined with a
synthetic spectrum for the object (namely a WD synthetic spectrum
or a WD+disk composite spectrum).

\section{Results}

Though the distance estimate to EM Cyg is most like around
400pc \citep{bai81,jam81,win03,wel05}, we do not fix 
the distance in our model fitting.  
We adopt a WD mass of $1M_{\odot}$ \citep{wel07} in all our models, with a
corresponding radius of $\sim 6,000$km (giving $Log(g) \sim 8.6$). All the
model fits are carried out on the first exposure (with the maximum flux).   
The reddening toward EM Cyg is $0.05\pm0.03$. 
As a consequence, we model EM Cyg assuming both E(B-V)=0.0 and E(B-V)=0.05. 
We first carry out the fitting without dereddening the spectrum,
and only later do we deredden the spectrum (assuming (E(B-V)=0.05)
to assess the effect of reddening on the results.

We first run single disk models. 
We find that in order to fit the short wavelengths of {\it FUSE},  
the best single disk model has  
a mass accretion rate $\dot{M} \sim 10^{-9}M_{\odot}$yr$^{-1}$, 
i.e. far larger than expected for quiescence, 
and $i\sim 60^{\circ}-75^{\circ}$.  
The $i=75^{\circ}$ model is shown in Figure 9. It gives a distance of 441pc
and $\chi^2_{\nu} = 0.6916$ (all our $\chi^2$ values are smaller
than one due to the low S/N of the {\it FUSE} exposures). 
On top of the model, we  add an ISM model in order to optimize the fit.  
The ISM has the following parameters  
$N_H =2 \times 10^{19}$ atoms/cm$^2$ , 
$N_{H_2} =1 \times 10^{17}$ molecules/cm$^2$,
T=160K, v=25km/s, $D/H=1\times 10^{-5}$, and   Z=0$\times$ solar.
We did not try to optimize the ISM model itself, nor to obtain accurate
parameter values for the ISM. The modeling the ISM is 
carried only in the context of improving the disk/WD model fit and to
help identify the  lines and spectral features. 
It is clear from Figure 9 that this disk model only vaguely follows
the continuum of the spectrum and does not fit the absorption features 
(except for the ISM lines). 
The two best single disk models are listed in Table 3.

Next, we run single WD models. 
We find that in order to fit the continuum the WD must have a temperature
of 40,000K, with a corresponding distance of 280pc . 
The $\chi^2$ obtained is $\chi^2_{\nu} \approx 0.58$. 
The two first WD models in Table 3 shows how the distance
and $\chi^2$ change as a function of the WD temperature. In this first
set we chose a guessed rotational velocity of 200km/s.    
In the next model (Table 3), we improve the $\chi^2$
value by varying the rotational velocity. The optimal
value is somewhere between $90$ and $100$km/s, and we decide to fix
the rotational velocity to 100km/s.  

In the next four models (Table 3) we change the abundances of silicon, sulphur and
nitrogen while keeping all the other abundances to their solar value
(we found that a solar carbon abundance is the best fit, though the
main carbon line at 1175\AA\  has also some broad emission and makes 
it difficult to model). 
The best fit is obtained for enhanced abundances as follows:
Sulphur: 10 times solar, Nitrogen: 10 time solar and Silicon: 20-30 times 
solar. We obtain a distance less than anticipated: $\sim$260pc. 
We present this best fit model in Figure 10. 
In order to optimize the fit we shift the synthetic WD spectrum to the
blue by 0.5\AA\ , to account for the radial velocity shift. This improves
slightly the $\chi^2$ values (the values listed in Table 3 reflect this
improvement). We also add an ISM model to the WD model, however,
the ISM model is not shifted.  

In the next model in Table 3 we deredden the spectrum assuming E(B-V)=0.05,
and obtain the same results, except for the distance. Here the distance is
barely 200pc.  We note that assuming
a WD mass of $\sim0.8M_{\odot}$ can increase the distance to 350pc without
changing the remaining parameters of the fit. On the other side, 
adding an accretion disk does increase the distance too, and this is
what we do next. 

For the WD+disk composite model, we first take our best single WD model and add a disk
model to it, to try and improve the fit and increase the distance. 
We find that the mass accretion rate has to be at least 
$\dot{M}=10^{-10}M_{\odot}$yr$^{-1}$ for the disk to contribute 
a noticeable flux. As the disk is added, the model deteriorates
and give a larger $\chi^2_{\nu}$.   We present such a model in Figure
11. The fit in the lower wavelength is not as good as the
best fit single WD model presented in Figure 10. 

In order to try
and improve the fit in the shorter wavelength we increase
the WD temperature,  as the mass accretion rate is already too large
for quiescence and therefore cannot be increased.   We find that
the fit  in the shorter wavelengths can be improved by elevating
the WD to a temperature $T\approx 50,000$K, however the overall
$\chi^2_{\nu}$ is further increased by doing so. This model is shown 
in Figure 12. While the model does well at fitting the overall continuum
of the spectrum, it does not fit the {\it FUSE} absorption lines as the velocity
broadening in the disk makes the overall depth of the line shallower
in the WD+disk model. It is possible that the absorption lines form
above the disk and WD; we discuss this possibility and its
implications in the next and last section.  

\section{Discussion and Conclusions} 

The over abundances and the unidentified lines in the {\it FUSE} spectrum  
of EM Cyg might be associated with some of the veiling material
rather than the WD itself, we have no additional way to verify this. 
However, 
simulations of the stream disk interaction  \citep{kun01} show that the
stream overflow hits the disk at smaller radii where it can in part 
bounce back. If this is the case, stream material can be found at any
orbital phase, but with greater density and altitude above the disk plane  
around $\phi \sim 0.6-0.9$. The rich variety of lines observed
in the present {\it FUSE} spectrum of EM Cyg is reminiscent of the 
{\it HST}/STIS  
spectrum of the DN TT Crt observed in a low state with an inclination
of $i\approx 60^{\circ}$ \citep{sio08}, and of the {\it FUSE} spectrum 
of the NL BB Dor 
possibly with an inclination of $80^{\circ}$ \citep{god08}. 
These spectra suggest line formation in 
different temperature regions, whether due to a wind above the disk 
(as could be the case in BB Dor) 
or overflowing stream material (possibly for TT Crt). 
So the strong and deep absorption lines in the {\it FUSE} exposure
1 of EM Cyg are likely due to circumstellar material, i.e. stream
material overflowing the disk rim and bouncing off the disk at
smaller radii, while the lower flux in exposures 2 and 3 are due to
the direct veiling of the WD by the stream overflow around phase 0.7.  

If the veiling material is responsible for the absorption lines, 
then the abundance fitting is not valid anymore, as the lines might be 
out of thermodynamic equilibrium or represent highly variable optical 
depths. As a consequence, 
the best fit to the continuum is actually the WD+disk model shown
in Figure 12, rather than the single WD model.
While the single WD model has a lower $\chi^2$ value, the 
WD+disk model is in better agreement with the distance of the system
and its higher-than-quiescent state. Because of this, 
we conclude that the {\it FUSE} spectrum of EM Cyg is best interpreted
as the spectrum of a hot WD with a small contribution from a low
mass accretion disk.  

The picture that emerges from this spectral analysis, is that 
of a 50,000K WD, contributing 93\% of the total FUV flux,
and a disk, with $\dot{M}=10^{-10}M_{\odot}$yr$^{-1}$ contributing 
the remaining 7\%. Such a composite WD+disk model gives a distance d=382pc in
best agreement with its expected value. The WD+disk model is also in
agreement with the moderately higher-than-quiescent state during
which the {\it FUSE} spectrum was taken, even though the $\chi^2$
value is slightly larger as the model does not fit the absorption
lines very well.  
The spectrum presents some deep absorption lines as well as
unidentified lines in the
longer wavelength region ($> 1100$\AA ), apparently forming in
front and above the WD/disk, from stream material overflowing
the disk edge. 
As a consequence, the exposure taken at phase $\phi \sim 0.45$ 
has a higher flux than
exposures obtained around $\phi \sim 0.65-0.90$. 
The large disk ($0.3a$) in EM Cyg is consistent with this picture, as 
the hot spot is consequently located closer to the secondary 
(around $\phi \sim 0.9$) and L1 stream material can more easily be
found off the disk plane. The large disk is also consistent with
the non-negligible mass accretion rate ($10^{-10}M_{\odot}$yr$^{-1}$).

However, whether modeled with a single WD component or a
WD+disk composite model, the {\it FUSE} spectrum of EM Cyg reveals a
rather hot WD. This result is in strong
disagreement with the previous spectral analysis of 
\citet{wel05,win03,urb06} who found a WD temperature
$T \sim 20-24,000$K. The discrepancy with previous analysis 
is due to (i) veiling of
the WD by stream material overflowing the disk rim (in both the
{\it FUSE} and {\it IUE} spectra around orbital phase $\phi \sim 0.6-0.9$,
which were modeled as such); 
(ii) contribution of the hot spot and secondary impact region to the longer 
wavelengths of the {\it IUE} SWP spectra; and (iii) eclipse of the
disk in the SWP 08088 {\it IUE} spectrum (taken around orbital phase
$\phi \sim 0.98-0.16$).  
In addition, in these previous analysis, the flux was contributed mostly from
an accretion disk, while in the present work the main component is the WD. 

Originally, the derivation of EM Cyg system parameters were affected by the 
presence of the third star, Arkadash, in the background of the system. 
In a similar way, we find that the previously derived mass accretion rate and 
WD temperature might have been affected by the veiling of the WD and disk by 
stream overflow around orbital phase $\phi \sim 0.65-0.90$.

\acknowledgments

PG wishes to thank Steve (Stephen) Lubow for some interesting 
discussions on the stream disk interaction, and  
Mario Livio for his kind hospitality at the 
Space Telescope Science Institute, 
where part of this work was carried out.  
We wish to thank an anonymous referee for her/his very prompt report
and constructive criticism. 
We are thankful to the members of the American Association of
Variable Star Observers (AAVSO)
for providing public online optical archival data on EM Cyg.         
This research was based on observations made with the
NASA-CNES-CSA Far Ultraviolet Spectroscopic Explorer.
{\it{FUSE}} is operated for NASA by the Johns Hopkins University under
NASA contract NAS5-32985.
Support for this work was provided by the National
Aeronautics and Space Administration (NASA) under Grant number 
NNX08AJ39G issued through the Office of Astrophysics Data Analysis
Program (ADP) to Villanova University (P.Godon).

\clearpage

\setlength{\hoffset}{-10mm} 
\begin{deluxetable}{ccl}
\tablewidth{0pc}
\tablecaption{System Parameters for EM Cyg}
\tablehead{Parameter & Value    & References  } 
\startdata 
$P_{orb}$    & 7hr/0.29090912d  &  \citet{mum69}   \\
$i$          & $67\pm 2 deg$    &  \citet{nor00}      \\ 
$M_{wd}$     & $1 M_{\odot}$    &  \citet{wel07,rob74,sto81} \\  
$M_{2nd}$    & $0.77 M_{\odot}$ &  \citet{wel07,rob74,sto81} \\  
$d$          & 350-500pc        &  \citet{bai81,jam81,wel05}  \\ 
$V_{min}$    & 14.4             &  \citet{spo03,spo05}  \\ 
$V_{max}$    & 12.5             &  \citet{spo03,spo05}  \\ 
DNO          & 14.6-16.5s       &  \citet{sti82}       \\ 
2nd Spec.Type & K3V             &  \citet{nor00}       \\  
Arkadash Spec.Type & K2V-K5V    &  \citet{nor00}      \\  
Gal.Lat.$~b$  & 4.277$^{\circ}$ &  SIMBAD             \\ 
E(B-V)        & 0.05$\pm$0.03   &  \citet{ver87,lad91,bru94} \\ 
\enddata 
\end{deluxetable} 

\clearpage 

\begin{deluxetable}{cccccccc}
\tablewidth{0pc}
\setlength{\tabcolsep}{0.04in} 
\tablecaption{FUSE and IUE Observations Log}
\tablehead{
Obs.  & Date & Time   &  Exp.time & Dataset & Start & Phase$^1$ & V  \\ 
          &(dd/mm/yy)& (UT) & (sec)$~~~$&  & HJD  & ($\phi$ ) & (range)   
}
\startdata
FUSE & 05/09/02 & 11h34m11s & 1,479 & C0100101001 & 2452522.9849 & 0.42-0.48 & 13.1-13.9 \\
FUSE & 05/09/02 & 12h54m33s & 2,651 & C0100101002 & 2452523.0407 & 0.62-0.73 & 13.1-13.9 \\
FUSE & 05/09/02 & 14h34m31s & 2,648 & C0100101003 & 2452523.1102 & 0.85-0.96 & 13.1-13.9 \\
FUSE & 05/09/02 & 16h22m59s & 2,135 & C0100101004 & 2452523.1855 & 0.11-0.19 & 13.1-13.9 \\
\hline 
IUE  & 29/02/80 & 14h42m50s & 2,700 & LWR 07056    & 2444299.1102 & 0.86-0.98 & 13.6-13.7 \\
IUE  & 29/02/80 & 15h32m22s & 4,500 & SWP 08088    & 2444299.1445 & 0.98-0.16 & 13.6-13.7 \\
IUE  & 05/08/82 & 02h40m47s & 1,500 & LWR 13862    & 2445186.6152 & 0.65-0.71 & 13.2-13.5 \\
IUE  & 05/08/82 & 03h11m30s & 1,500 & SWP 17591    & 2445186.6366 & 0.72-0.78 & 13.2-13.5 \\
IUE  & 05/08/82 & 03h44m26s & 2,400 & LWR 13863    & 2445186.6595 & 0.81-0.91 & 13.2-13.5 \\
IUE  & 05/08/82 & 04h34m08s & 2,399 & SWP 17592    & 2445186.6940 & 0.92-0.02 & 13.2-13.5 \\
\enddata 
\tablenotetext{1}{The ephemeris, time of minima/eclipse HJD2400000 + 43780.7508, 45257.4091,
50692.4613, \& 53709.1873 were taken from \citet{mum80,beu84,csi08} with the period
0.29090912d=25134.55s} 
\end{deluxetable}

\clearpage 

\setlength{\hoffset}{-15mm} 
\begin{deluxetable}{cccccccccccc}
\tablewidth{0pc}
\tablecaption{WD Synthetic Spectra}
\tablehead{
$T_{wd}$ & $V_{rot}sin{i}$& [Si] & [S] & [N] & $\dot{M}$&$i$&d&$\chi^2_{\nu}$&WD/Disk&$E_{B-V}$ & Figure\\
($10^3$K)&       (km/s)  &(Solar)&(Solar)&(Solar)&($M_{\odot}$yr$^{-1}$)&(deg)&(pc)& & \%  &  \\
}
\startdata
 ---       &  ---  & 1.0   &  1.0  &  1.0  & $10^{-9}$ & 75 & 441  &  0.6916 & 0/100  & 0.00   & 9  \\  
 ---       &  ---  & 1.0   &  1.0  &  1.0  & $10^{-9}$ & 60 & 754  &  0.6867 & 0/100  & 0.00   &    \\  
 35        &  200  & 1.0   &  1.0  &  1.0  &     -     & -  & 223  &  0.6063 & 100/0  & 0.00   &     \\ 
 40        &  200  & 1.0   &  1.0  &  1.0  &     -     & -  & 280  &  0.5823 & 100/0  & 0.00   &    \\ 
 40        &  100  & 1.0   &  1.0  &  1.0  &     -     & -  & 280  &  0.5709 & 100/0  & 0.00   &     \\ 
 40        &  100  & 30.   &  10.  &  1.0  &     -     & -  & 264  &  0.4780 & 100/0  & 0.00   &    \\ 
 40        &  100  & 10.   &  10.  &  10.  &     -     & -  & 267  &  0.4847 & 100/0  & 0.00   &    \\ 
 40        &  100  & 20.   &  10.  &  10.  &     -     & -  & 264  &  0.4768 & 100/0  & 0.00   &   \\ 
 40        &  100  & 30.   &  10.  &  10.  &     -     & -  & 261  &  0.4767 & 100/0  & 0.00   & 10 \\  
 40        &  100  & 30.   &  10.  &  10.  &     -     & -  & 197  &  0.4776 & 100/0  & 0.05   &    \\ 
 40        &  100  & 30.   &  10.  &  10.  & $10^{-10}$& 60 & 318  &  0.5291 &  77/23 & 0.00   & 11 \\  
 50        &  100  & 30.   &  10.  &  10.  & $10^{-10}$& 60 & 382  &  0.5573 &  93/7  & 0.00   & 12 \\  
\enddata
\end{deluxetable}

\clearpage 

\setlength{\hoffset}{00mm} 

\clearpage 
\begin{figure}
\plotone{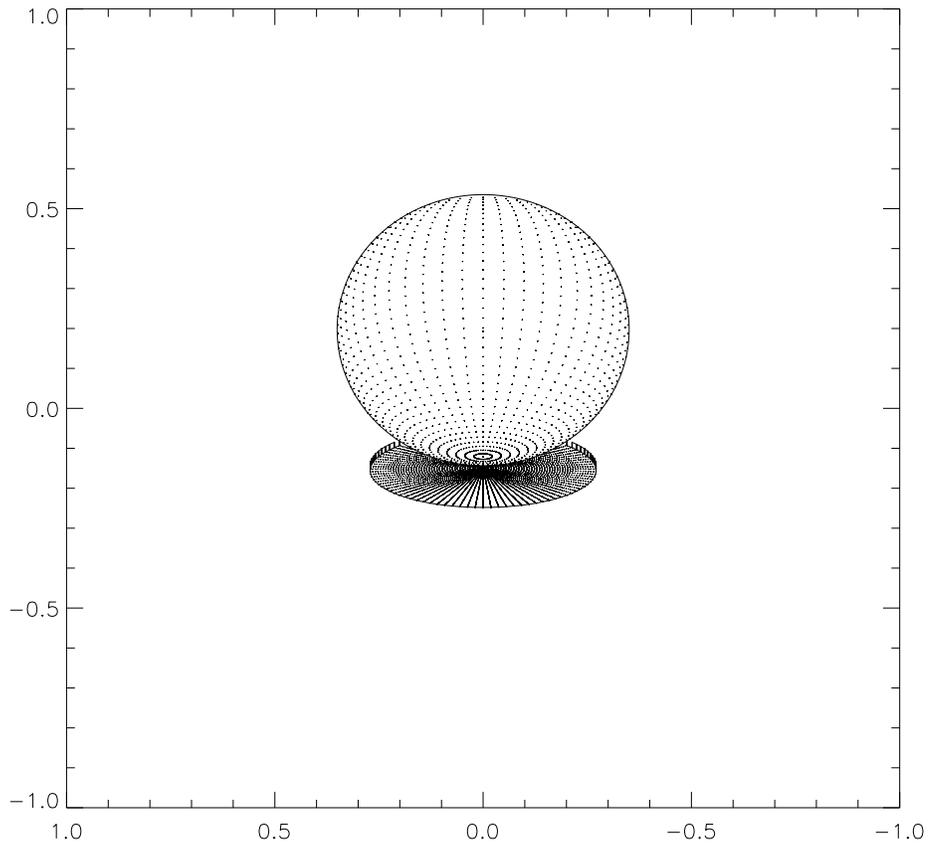}
\caption{The binary configuration of EM Cyg is shown during 
eclipse of the disk at orbital phase $\phi=0$. 
The inclination is $i=67^{\circ}$ and the
mass ratio is 0.77. The disk is eclipsed but the white dwarf
is not.} 
\end{figure}

\begin{figure}
\plotone{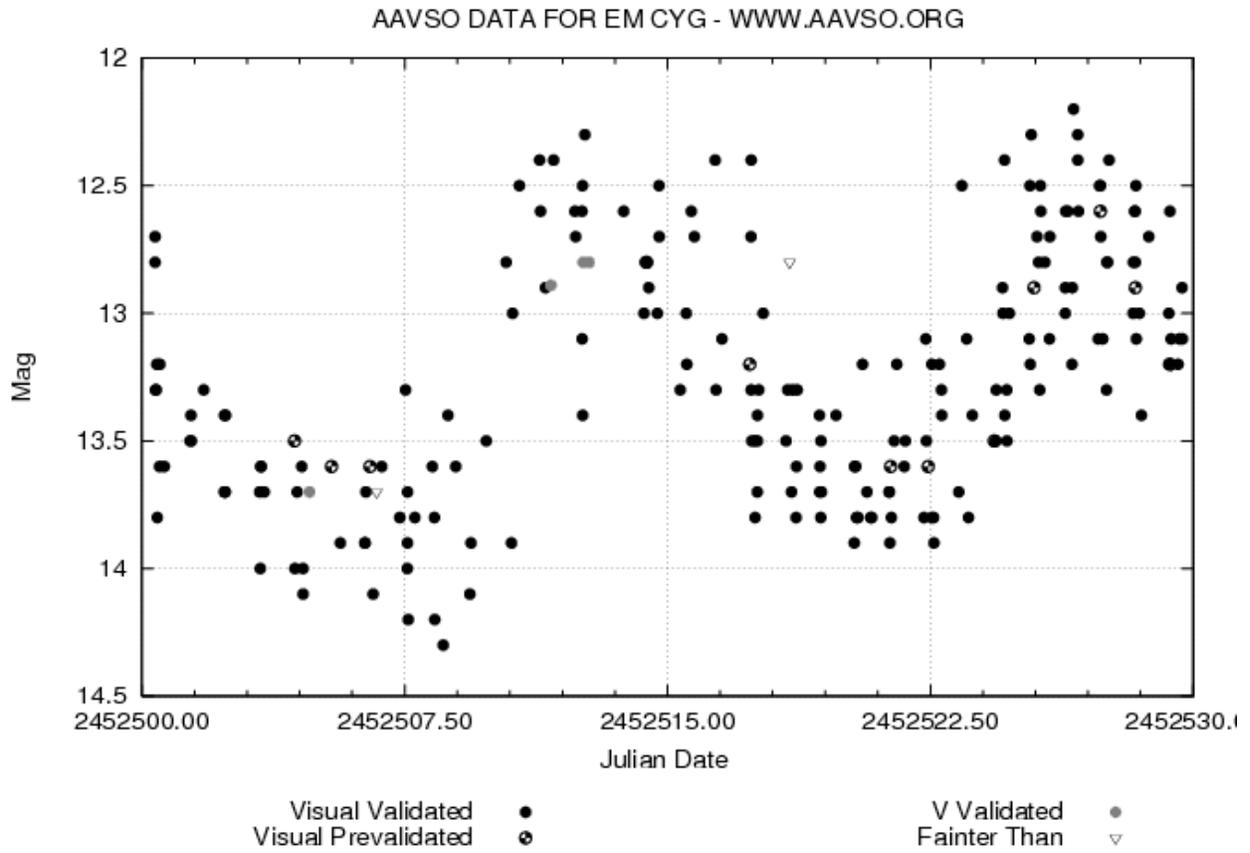} 
\figcaption{The visual magnitude of EM Cyg recorded by the members of the 
AAVSO (www.aavso.org).  
EM Cyg was observed with STIS (around JD2452515)  
when the system was in outburst (center of the graph),
and was observed with {\it FUSE} a week later (on JD2452523)
when the system was in quiescence.     
}  
\end{figure}

\clearpage 
\begin{figure}
\vspace{-6.cm} 
\plotone{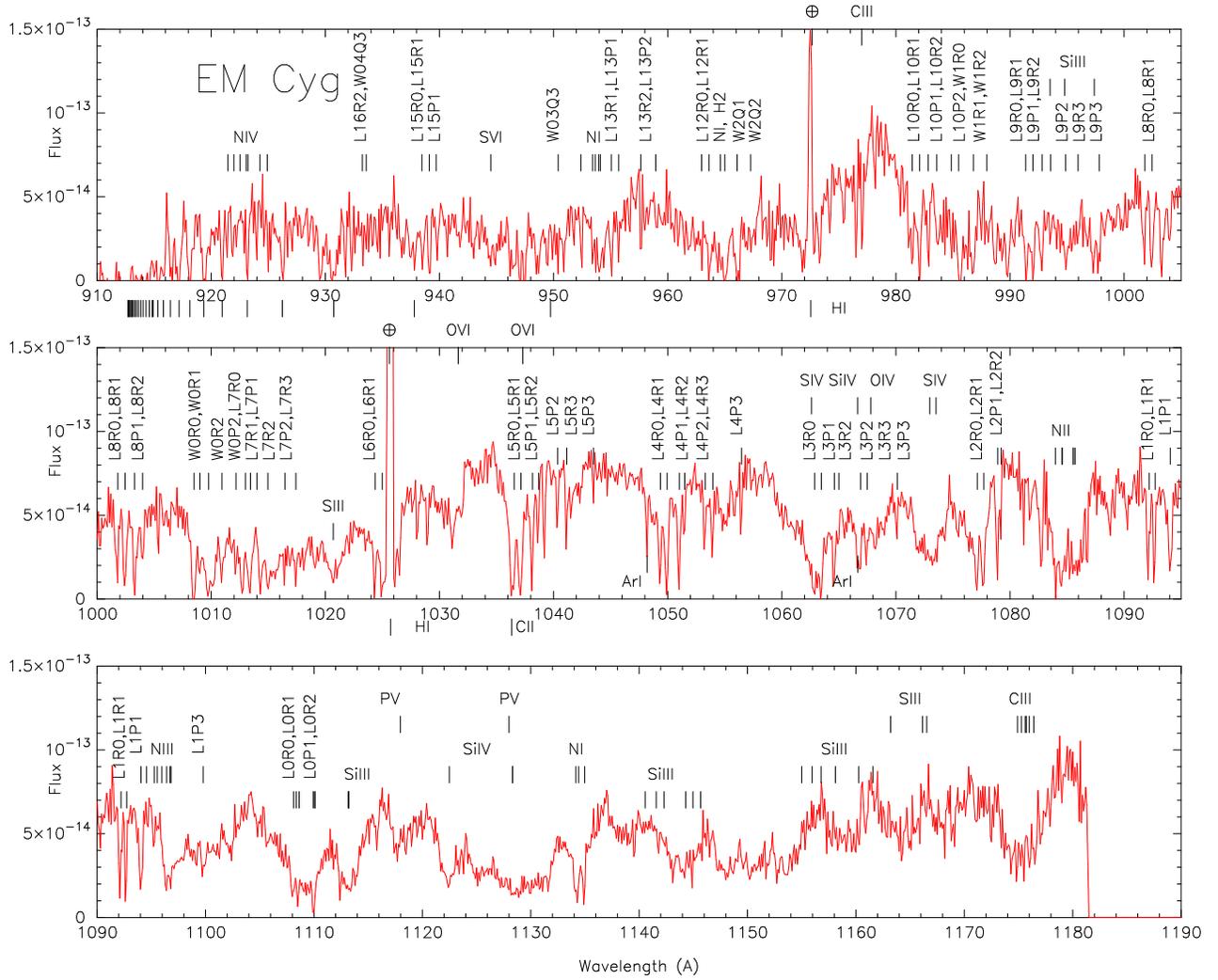} 
\vspace{+1.cm} 
\caption{The {\it FUSE} spectrum of EM Cyg. This spectrum consists of
4 exposures (corresponding to 4 FUSE orbits). The sharp emission lines
are contamination. The ISM molecular features are marked vertically for
clarity. The molecular lines are identified by their band (W=Werner;
L=Lyman), upper vibrational level (1-16), and rotational transition
(R, P, or Q with lower rotational state J=1-3). The ISM atomic lines
are marked below the x-axis. Some ISM metal lines are also identified:
N\,{\sc i}, Ar\,{\sc i}. The Nitrogen lines are possibly also contaminated
with terrestrial atmospheric absorption (especially the N\,{\sc i} 1135\AA ).
We have marked the position of the N\,{\sc iv} 924 \AA\ lines, though 
they are not detected. The C\,{\sc iii} and O\,{\sc vi} lines are in
broad emission, with a possible P-Cygni profile at the C\,{\sc iii} 1175\AA\ .}
\end{figure} 

\clearpage 
\begin{figure}
\plotone{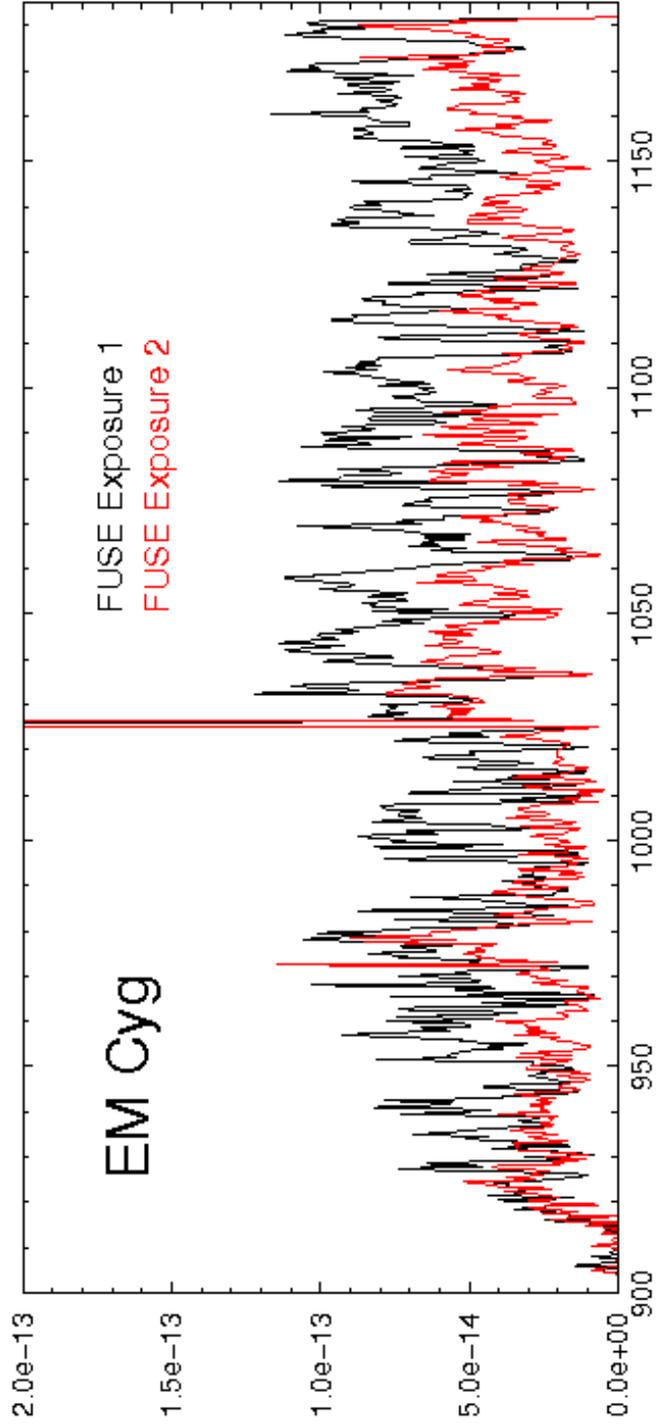} 
\caption{The {\it FUSE} exposures 1 \& 2 of EM Cyg. During the first exposure
the flux is about 2 to 4 times larger than during exposure 2. Exposure
3 is similar to exposure 2; exposure 4 is similar to exposure 1. 
For clarity exposures 3 and 4 are not shown, and the spectra have 
been binned at 0.5\AA .  
}
\end{figure} 

\clearpage 
\begin{figure}
\vspace{-1.cm} 
\plottwo{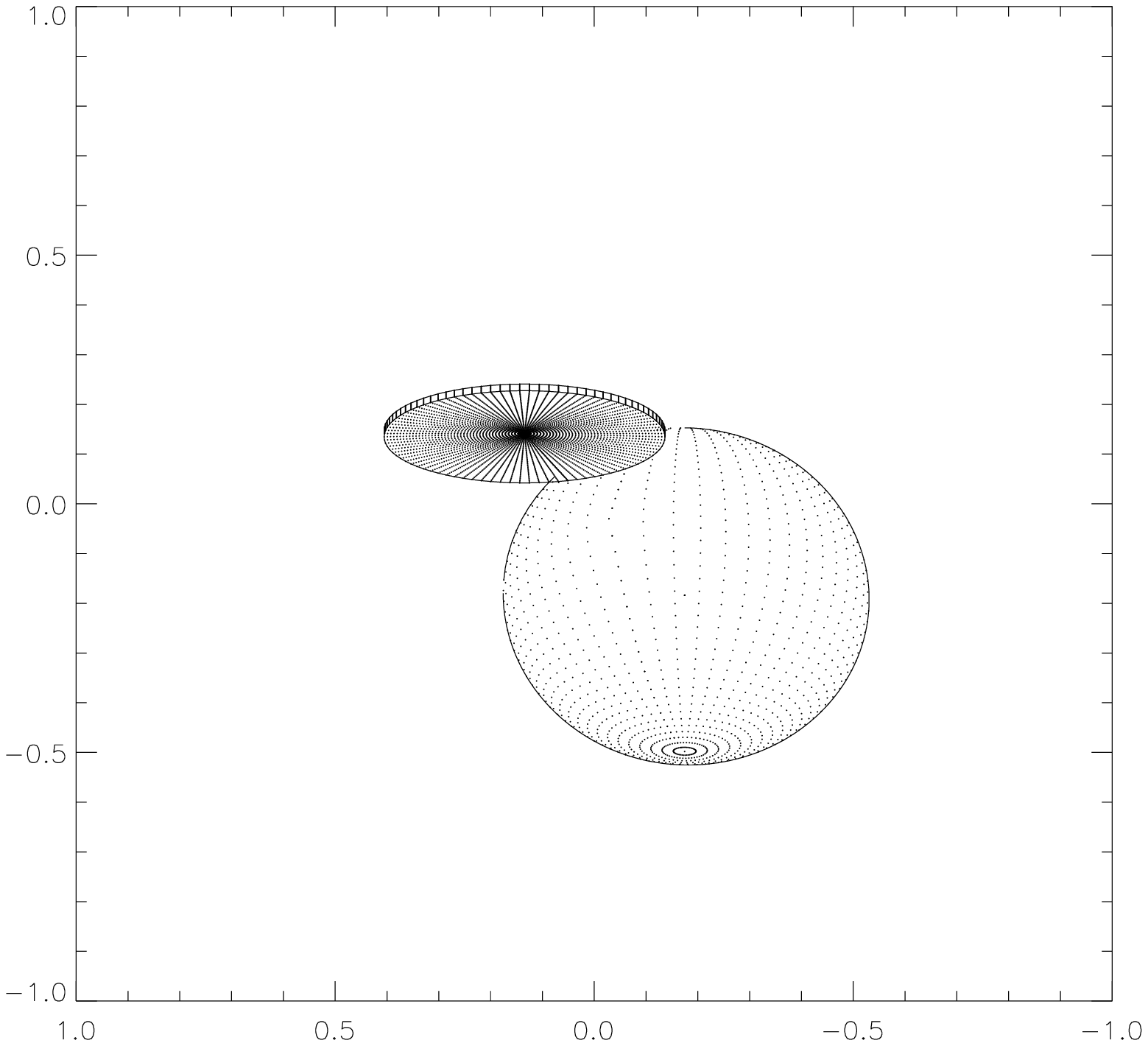}{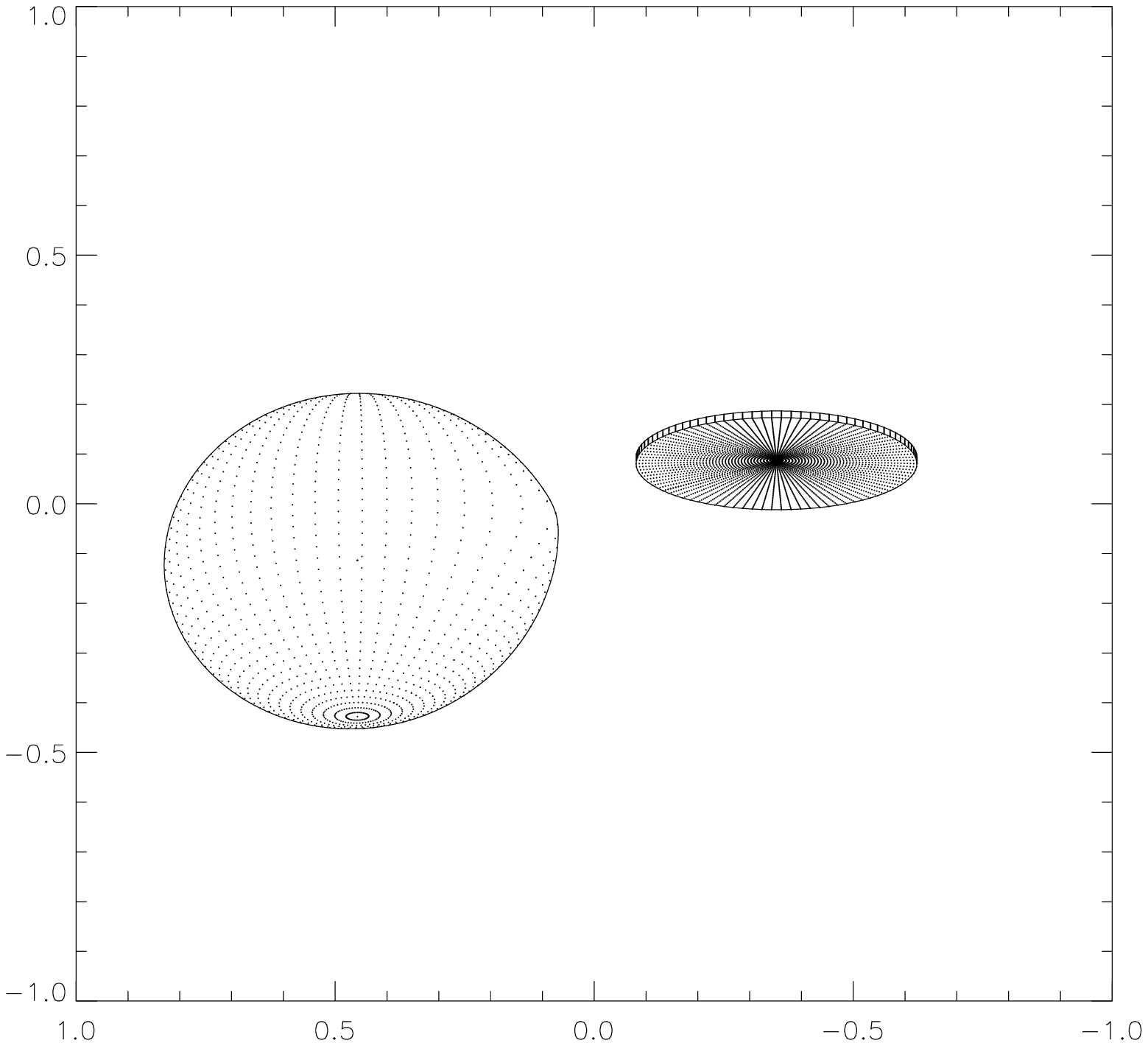} 
\end{figure} 
\begin{figure}
\vspace{-7.cm} 
\plottwo{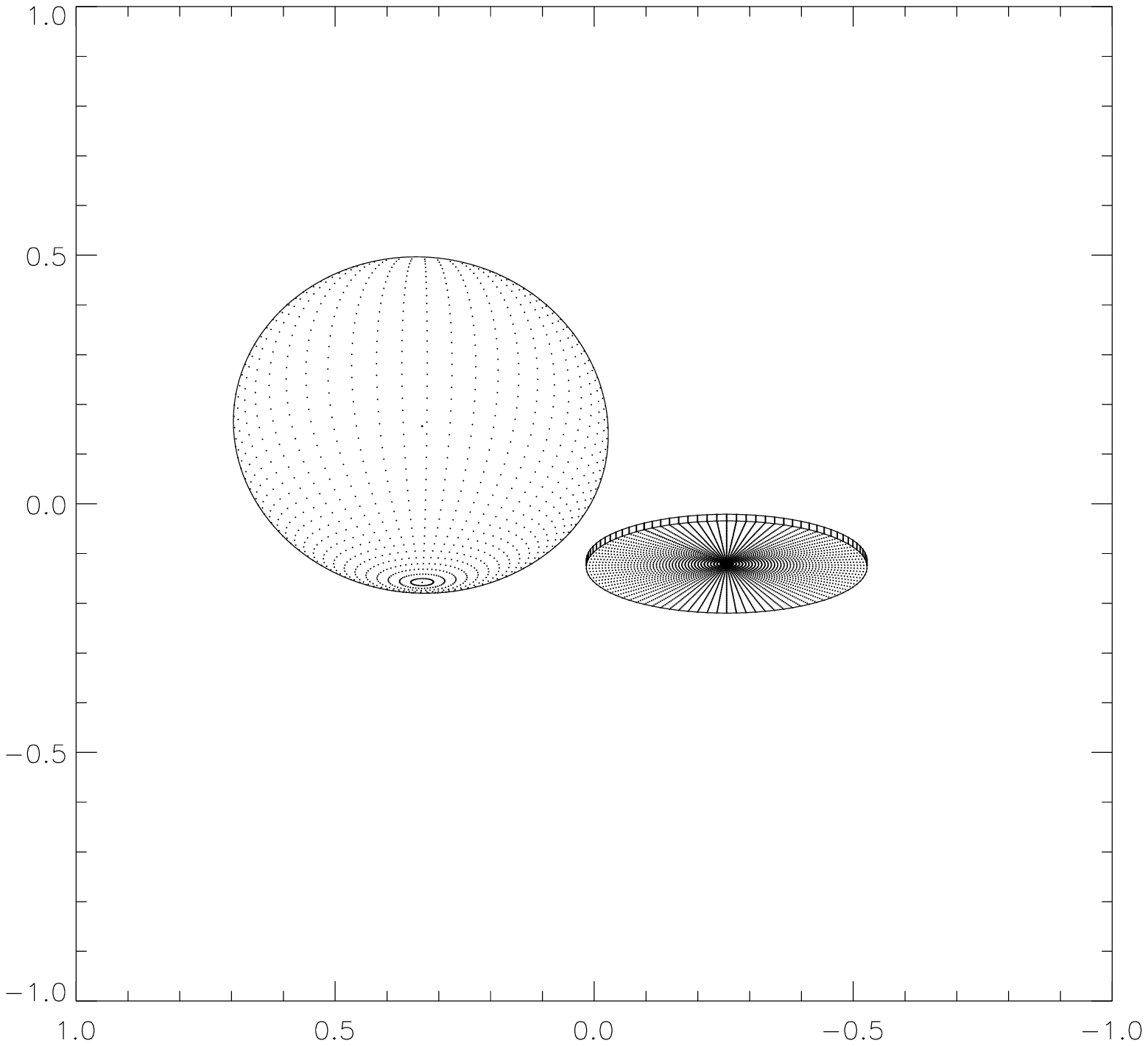}{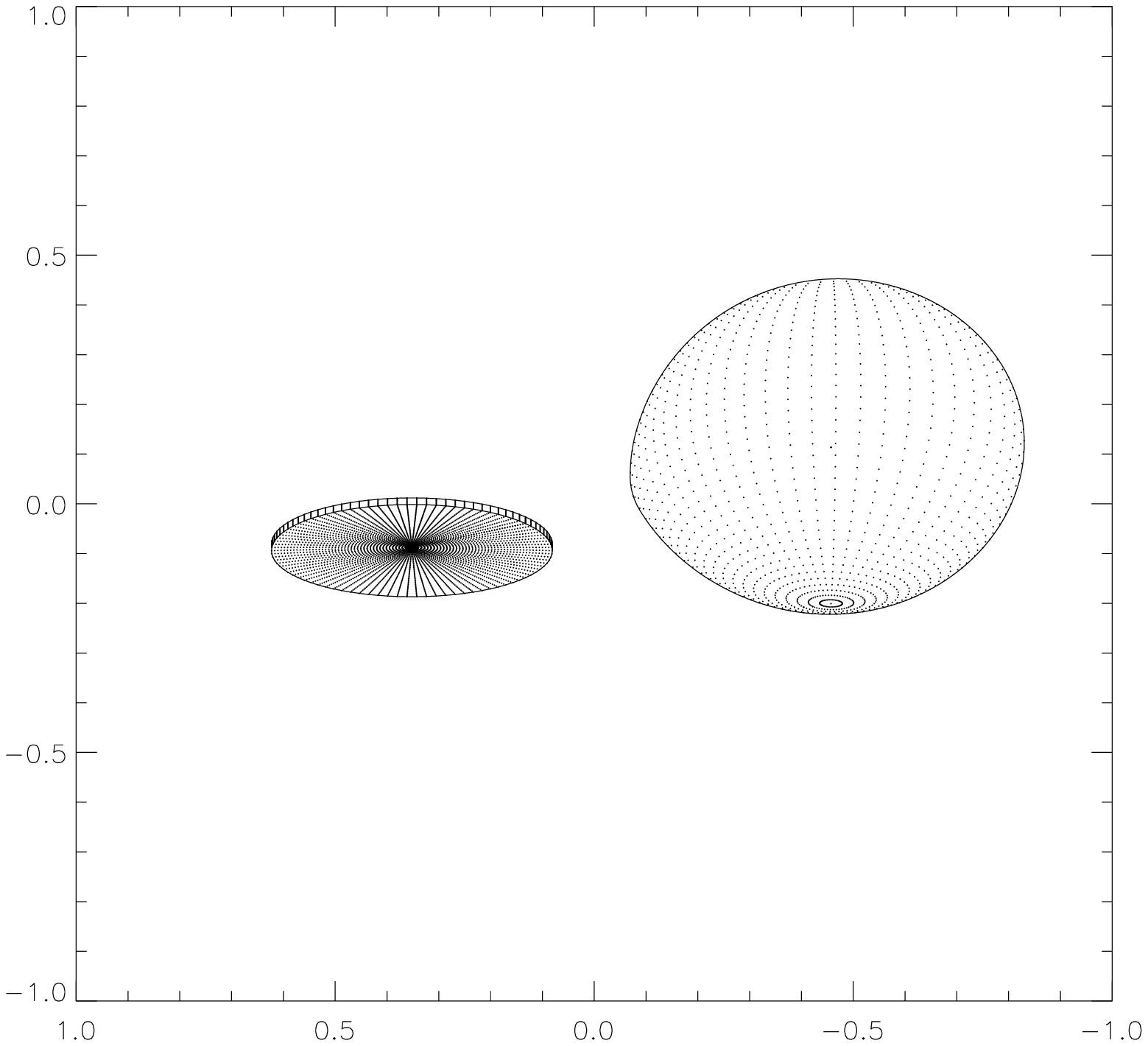} 
\caption{
EM Cyg Binary Configuration during the 4 FUSE orbits/exposures. 
Exp.1 (upper left) is at phase $\phi \sim 0.45$: 
the disk is in front of the observer and absorption is minimal.  
Exp.2 (upper right) is at phase $\phi \sim 0.65$: 
the disk is still slightly in front moving away from the observer, 
the stream overflow is in front, absorption is maximal. 
Exp.3 (lower left) is at phase $\phi \sim 0.90$: 
the secondary filling its Roche lobe is in front moving to the right, 
the stream and hot spot are directly in front, absorption is maximal.  
Exp.4 (lower right) is at phase $\phi \sim 0.15$: 
the secondary Roche lobe is slightly in front and to the right of the observer, 
the disk is in the back moving towards the observer.
The stream is in the back and behind, absorption is minimal.
}
\end{figure}

\clearpage
\begin{figure}  
\vspace{-1.cm} 
\plotone{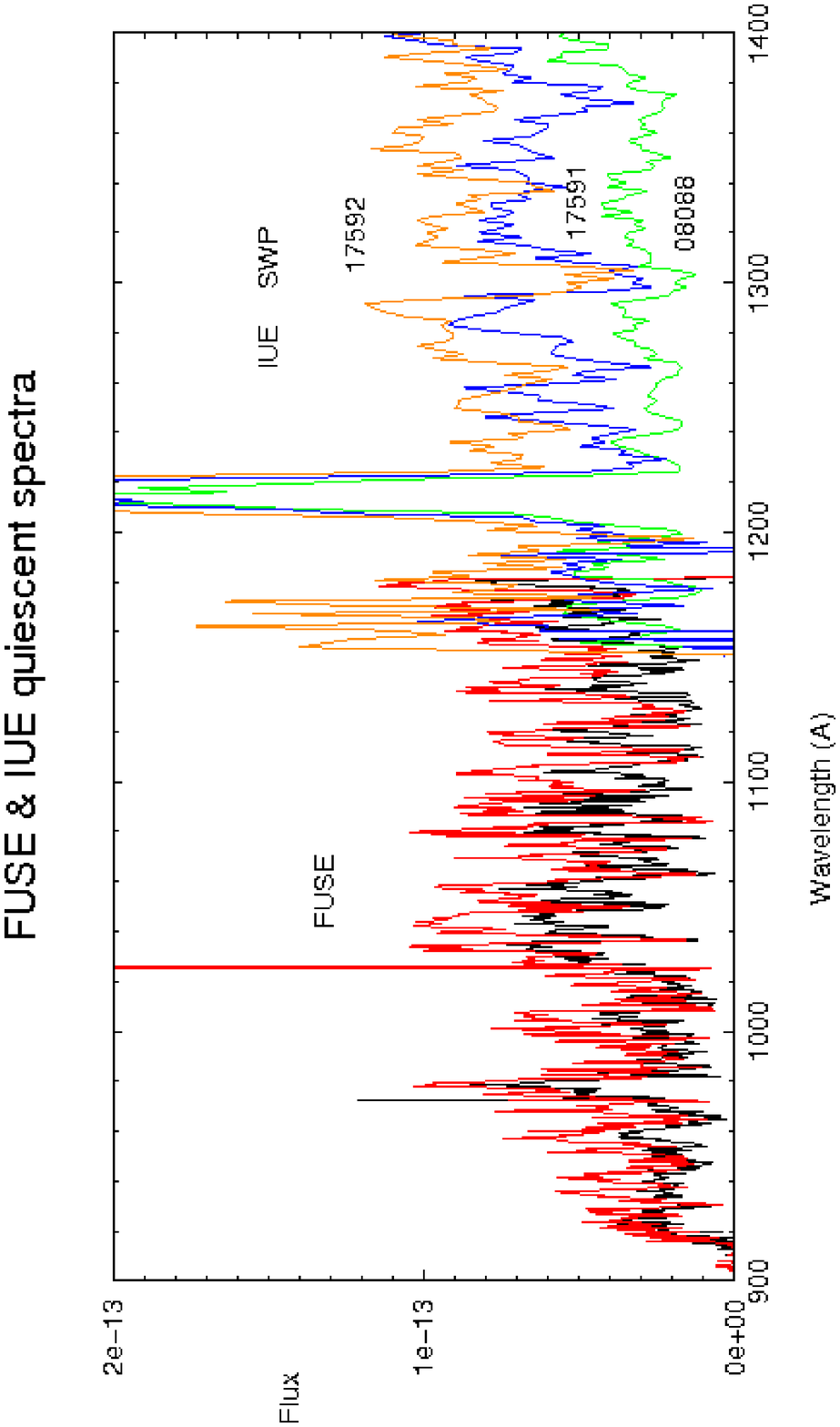} 
\vspace{-1.cm} 
\figcaption{The {\it FUSE} exposures are shown together with the
SWP {\it IUE} spectra. The {\it FUSE} exposures 1 and 4 have been 
combined together and are shown in red; exposures 2 and 3 have also
been combined together and are shown in black. {\it FUSE} exposures  
2 and 3 have a lower flux due to veiling of the WD by stream material
overflowing the disk. It is likely that the lower flux in the 
{\it IUE} spectrum SWP 08088 is also due to the veiling of the WD. 
Additional change in the {\it IUE} spectra might be due to the
cooling of the WD or ongoing accretion.}
\end{figure}

\clearpage
\begin{figure}  
\vspace{-1.cm} 
\plotone{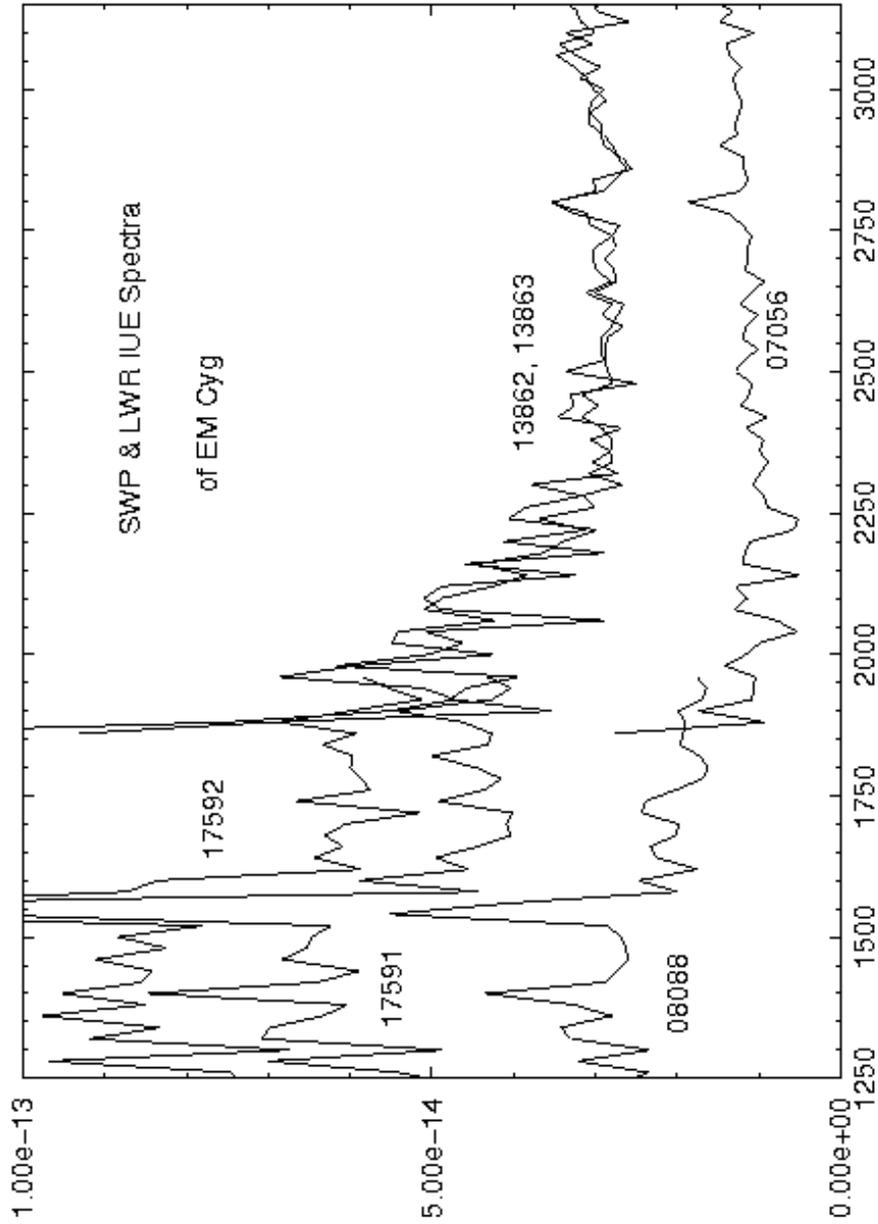} 
\vspace{-1.cm} 
\figcaption{The {\it IUE} LWR and SWP spectra of EM Cyg.
The wavelength is in \AA , and the flux is in 
ergs$~$s$^{-1}$cm$^{-2}$\AA$^{-1}$. The plateau             
between 2250 and 3000 \AA\ is possibly due to the hot spot
with $T\sim 10,000$K, as pointed out by \citet{szk81}. }  
\end{figure}

\clearpage 
\begin{figure}
\vspace{-5.cm} 
\plotone{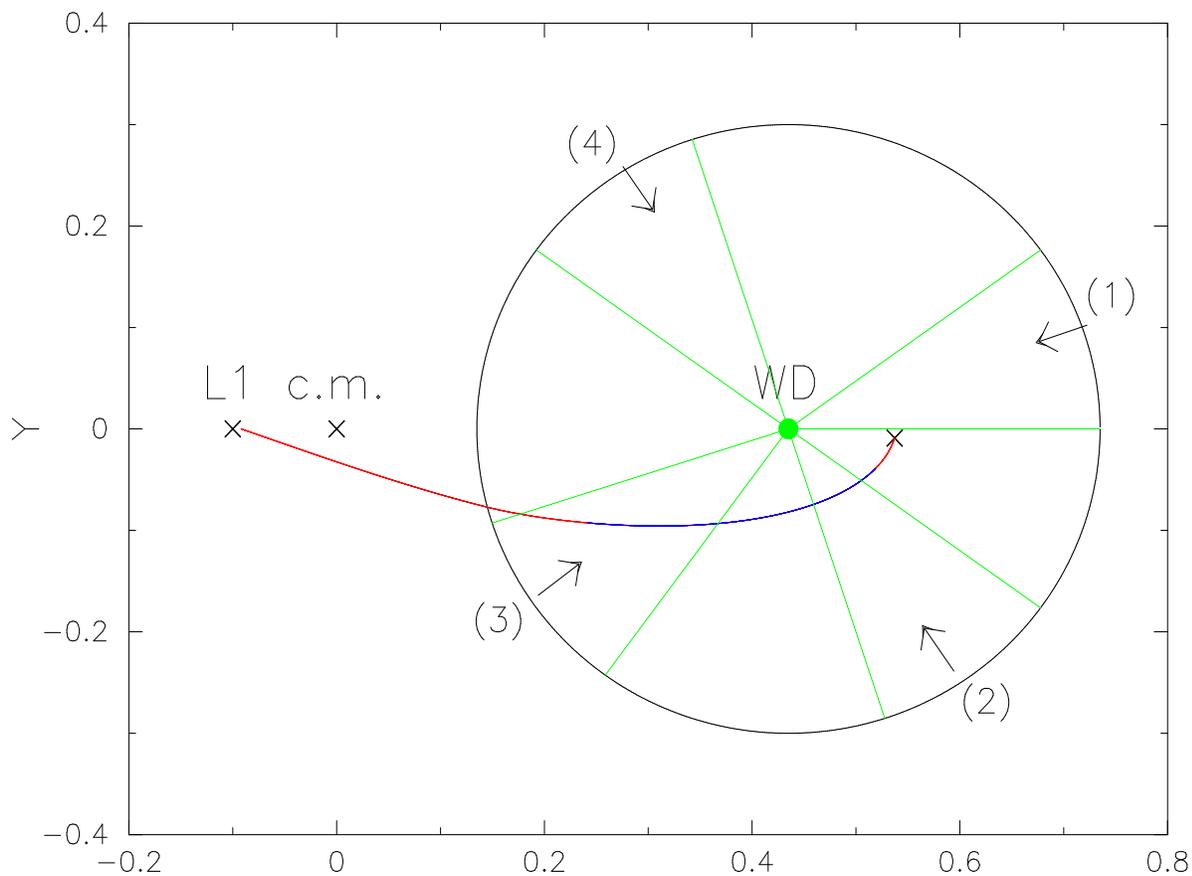} 
\vspace{-3.cm} 
\caption{
EM Cyg Binary Configuration during the 4 {\it FUSE} orbits/exposures. 
The areas delimited in green represent the viewing angles during 
the 4 {\it FUSE} exposures (1,2,3 \& 4).  
The disk radius here is $0.3a$, where $a$ is the binary separation. 
The stream trajectory from the L1 point is shown in red. The stream  
overflows the disk rim and from there it is launched at an angle
$\sim 55^{\circ}$ 
above the disk; it falls back onto the disk around $\phi \sim 0.5$.
The region for which the stream can reach $z/r \sim 0.4$ (corresponding
to veiling at $i\sim67$deg) is shown in blue (from \citet{god09}). 
} 
\end{figure}

\clearpage

\begin{figure}
\vspace{-5.cm} 
\plotone{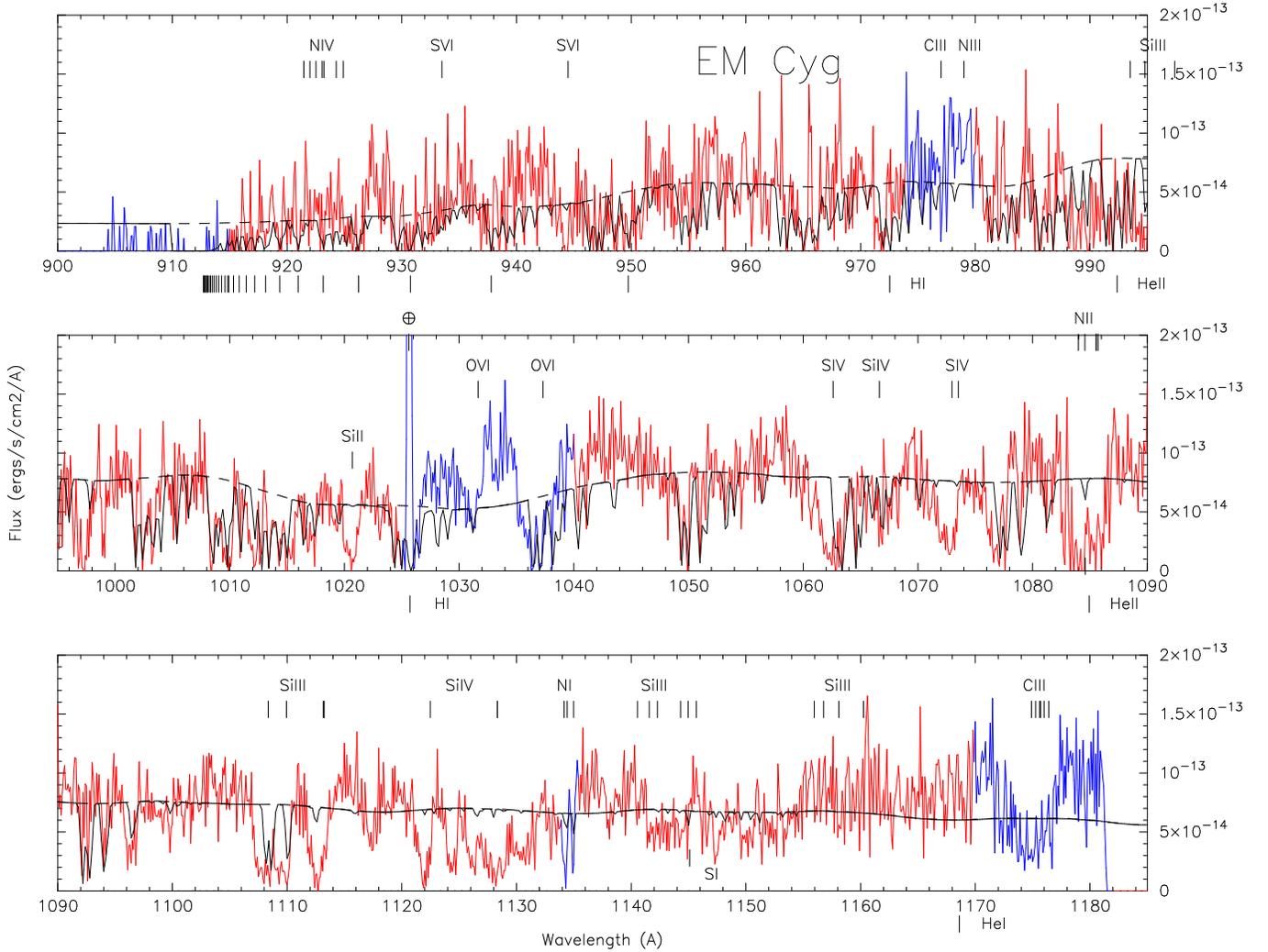} 
\caption{The best fit single disk model (in solid black) to the 
{\it FUSE}/exposure 1 spectrum of EM Cyg. The observed spectrum
(in red) has not been dereddened. 
The regions that have been masked for the fitting are shown
in blue. 
The model consist of 
an accretion disk around a $1 M_{\odot}$ WD with 
$\dot{M}=1 \times 10^{-9}M_{\odot}$yr$^{-1}$, an
inclination $i=75^{\circ}$. The distance obtained is
$d=441$pc and $\chi^2_{\nu} =0.6916$.  
Because of the rotational broadening the model cannot
fit the absorption lines from the source, and the mass accretion
rate is too large and inconsistent with the quiescent
state of the system.
An ISM model has been added to the disk model to fit the ISM lines.
The dotted line show the disk model without the inclusion of the
ISM model.} 
\end{figure}

\clearpage 
\begin{figure}
\vspace{-5.cm} 
\plotone{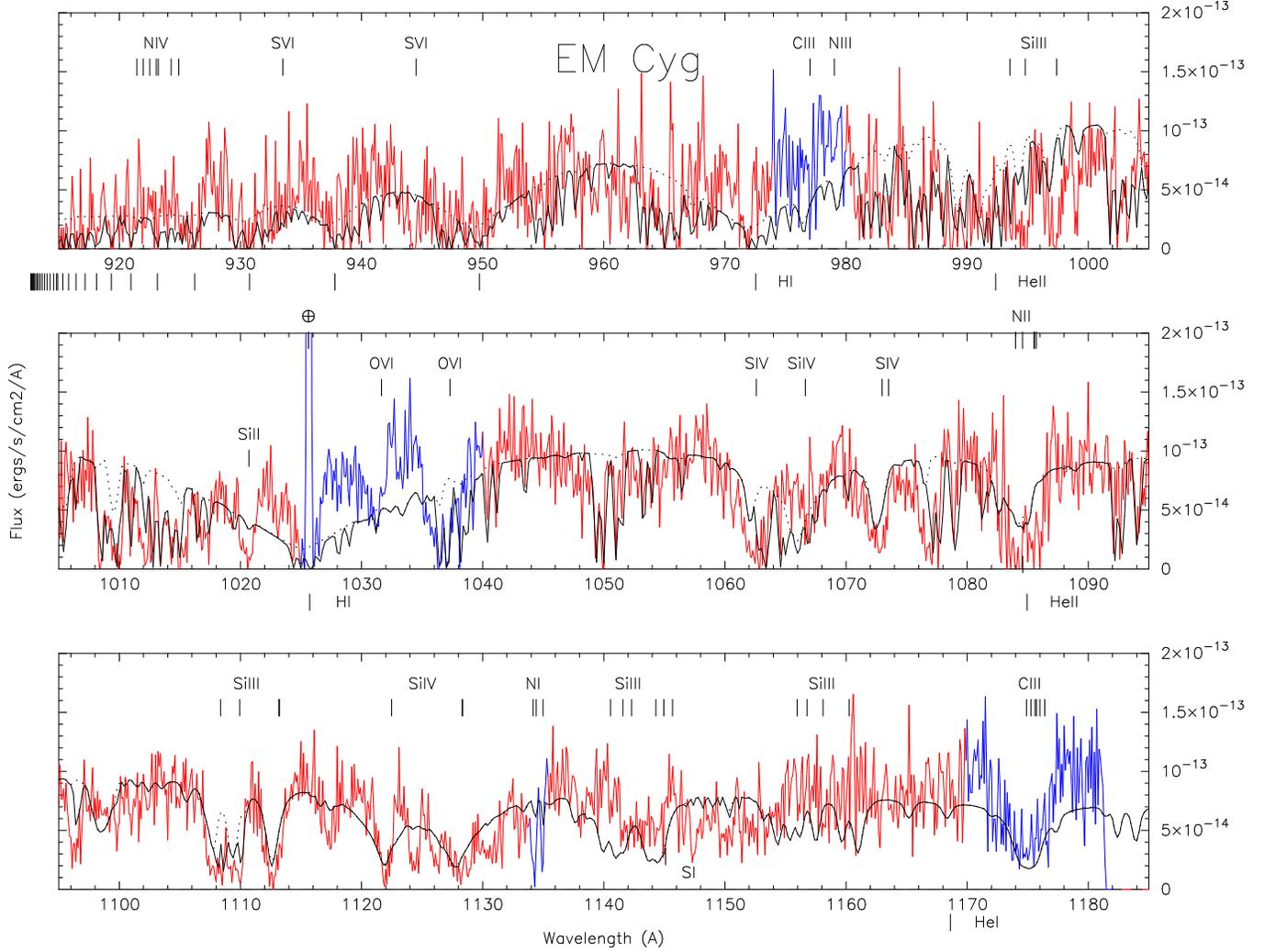} 
\vspace{-3.cm} 
\caption{The best fit WD model (in solid black) to the 
{\it FUSE}/exposure 1 spectrum of EM Cyg. The observed spectrum (in red)
has not been dereddened. The model consist of
a $1.0M_{\odot}$ WD, with $T=40,000$K, a projected rotational
velocity $V_{rot} \sin{i} =100$km/s, and over-solar abundances
of Si (30), S (10), and N (10). The entire WD synthetic spectrum   
has been shifted to the blue by $\sim0.5$\AA, which corresponds
to a radial velocity of about $\sim 150$km/s (at 1,000\AA )
in rough agreement with the radial velocity $+125$km/s \citet{wel05}. 
An ISM model (no shifted) has been added to improve the overall fit. 
The dotted line represents
the WD model without the inclusion of the ISM model. The regions of the
observed spectrum that have been masked for the fitting are shown in
blue. The distance obtained from the best fit is $d=261$pc and
$\chi^2_{\nu}=0.4768$  
}
\end{figure}

\clearpage
\begin{figure}
\vspace{-5.cm}
\plotone{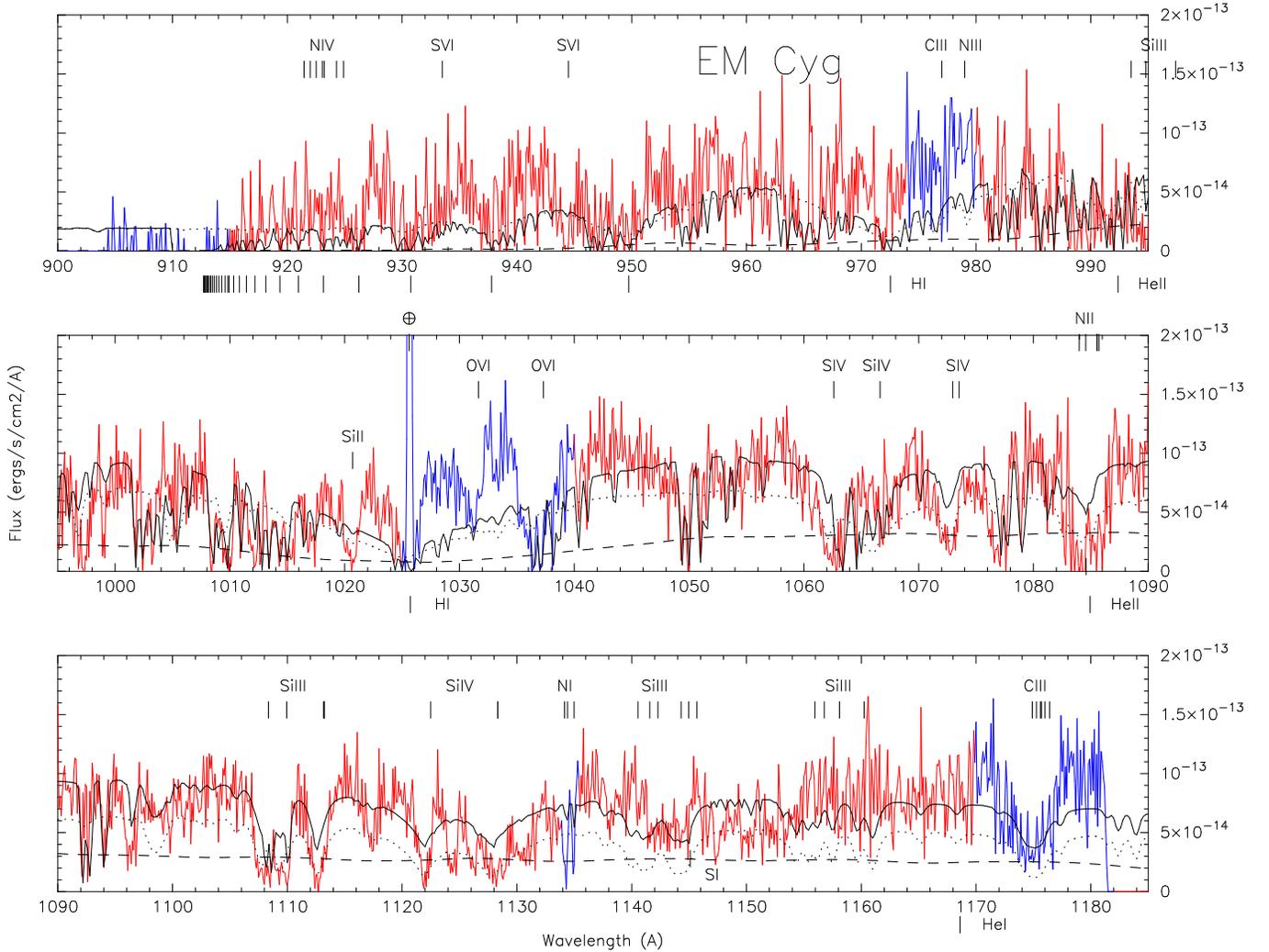} 
\caption{The best WD model (from Figure 10) plus a disk model 
are fitted (in solid black) to the 
{\it FUSE}/exposure 1 spectrum of EM Cyg. The observed spectrum
(in red) has not been dereddened. 
The regions that have been masked for the fitting are shown
in blue. 
The disk model (dashed line) consist of 
an accretion disk around a $1 M_{\odot}$ WD with 
$\dot{M}=1 \times 10^{-10}M_{\odot}$yr$^{-1}$, an
inclination $i=60^{\circ}$.
The WD model (dotted line) consists of
a $1.0M_{\odot}$ WD, with $T=40,000$K, a projected rotational
velocity $V_{rot} \sin{i} =100$km/s, and over-solar abundances
of Si (30), S (10), and N (10). The entire WD+disk synthetic spectrum   
has been shifted to the blue by $\sim0.5$\AA\ . 
The distance obtained is $d=318$pc and $\chi^2_{\nu} =0.5291$.  
An ISM model has been added to the WD+disk model to fit the ISM lines.
The WD contributes 77\% of the FUV flux and the disk contributes
the remaining 23\%. 
}
\end{figure}

\clearpage
\begin{figure}
\vspace{-5.cm}
\plotone{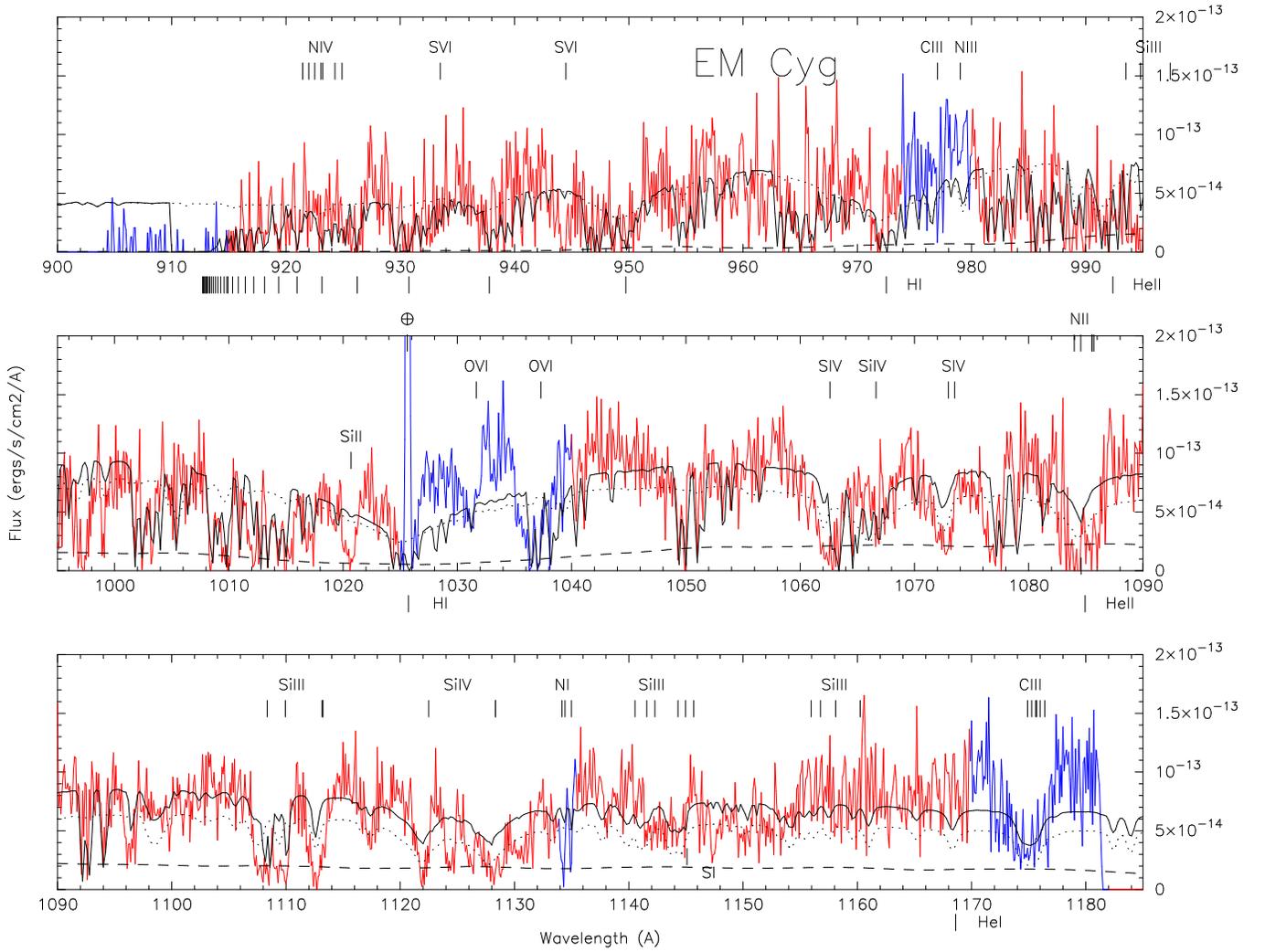} 
\caption{Same as in Figure 11, but here the WD temperature has
been increased to 50,000K. The WD contributes to 93\% of the FUV
flux and the disk contributes the remaining 7\%. 
The distance obtained is $d=382$pc and $\chi^2_{\nu}=0.5573$. 
This model fits the shorter wavelength continuum flux better 
than the previous models, but it does not fit the absorption 
features as well as the single WD model.  
}
\end{figure}

\end{document}